\documentclass[letter, amsfonts, amssymb, amsmath, reprint, showkeys, 
twoside, twocolumn,
elsarticle,
superscriptaddress]{revtex4-1}

\usepackage{xcolor}
\usepackage{graphicx} 
\usepackage{subfigure}
\usepackage{dcolumn} 
\usepackage{bm}
\usepackage{float}
\usepackage[columnwise,running]{lineno}
\usepackage[normalem]{ulem}

\usepackage{slashed}
\usepackage{verbatim}

\usepackage{hyperref}
\usepackage{setspace}
\usepackage{color}
 \usepackage{array, diagbox}
\definecolor{orange}{cmyk}{0.,0.353,1.,0.}    

\usepackage{parskip}
\usepackage{tikz}
\usetikzlibrary{positioning}

\begin{document}

\title{Search for the chiral magnetic effect through beam energy dependence of charge separation using event shape selection}

\author{The STAR Collaboration}

\begin{abstract}

High-energy, heavy-ion collisions can create local domains of chirality-imbalanced quarks, reflecting the topological features of quantum chromodynamics. 
The chiral magnetic effect (CME) predicts an electric charge separation of quarks in such topological domains along the magnetic field ($\vec{B}$) generated by the passing of two high-$Z$ nuclei.
We use a correlation observable $\Delta\gamma^{112}$ between charged meson pairs to detect the CME-induced charge separation and a novel event shape selection (ESS) method to mitigate the background effects related to elliptic flow ($v_2$). 
The ESS method classifies events based on the emission pattern of final-state particles and determines $\Delta\gamma^{112}_{\rm ESS}$ from the zero-flow limit. 
We reconstruct the $\vec{B}$ field direction from the spectator nucleons, which minimizes backgrounds unrelated to the collective motion of the system. 
In this work, we report the measurements of $\Delta\gamma^{112}$ and a background indicator $\Delta\gamma^{132}$ in Au+Au collisions from the RHIC Beam Energy Scan phase II and at the top RHIC energy. 
After background suppression, $\Delta\gamma^{132}_{\rm ESS}$ aligns with zero, and $\Delta\gamma^{112}_{\rm ESS}$ is reduced to no more than 20\% of $\Delta\gamma^{112}$. 
We observe a finite residual charge separation with $2.6\sigma$, $3.1\sigma$, and $3.3\sigma$ significance in the 20\%--50\% centrality range of Au+Au collisions at 11.5, 14.6, and 19.6 GeV.
The results at 17.3 and 27 GeV also show positive values but with a lower significance of $1.3\sigma$ and $1.1\sigma$, respectively. 
The corresponding $\Delta\gamma^{112}_{\rm ESS}$ values at 7.7, 9.2, and 200 GeV are consistent with zero within uncertainties.

\begin{description}
\item[keywords]
heavy-ion collision; chiral magnetic effect; event shape
\end{description}
\end{abstract}
\maketitle

\section{Introduction} \label{intro}

The finite mass of quarks in quantum chromodynamics (QCD) causes mixing between left- and right-chiral components, preventing quarks from maintaining a fixed chirality.
At the BNL Relativistic Heavy Ion Collider (RHIC) and the CERN Large Hadron Collider (LHC), the created quark-gluon plasma (QGP) may restore chiral symmetry, whereby deconfined light quarks become nearly massless and obtain a definite chirality.
In this state, a finite chiral chemical potential ($\mu_5\neq0$) can emerge in local QGP domains where topological vacuum transitions in excited gluon fields stimulate the chiral anomaly\cite{CME-1, CME-2, CME-3, Kharzeev:2024zzm}.
These chirally imbalanced local domains can be explored through the chiral magnetic effect (CME), which predicts an electric current ($\vec{J}$) due to the motion of the quarks inside the domains in the presence of an intense external magnetic field ($\vec{B}$), with $\vec{J} \propto \mu_5\vec{B}$~\cite{CME-1, CME-2, CME-3,Baryogenesis}.
As a precondition, a strong magnetic field on the order of $10^{18}$ Gauss is generated in noncentral ultrarelativistic collisions by highly charged ions passing by each other at nearly the speed of light~\cite{CME-5,B-field-1,B-field-2,STAR-Bfield}.
Thus, the detection of the CME in heavy-ion collisions provides a distinct channel to understand the non-trivial topological sector of QCD and the evolution of the magnetic field. 

For the past two decades, the search for CME-induced charge separation has been actively pursued in experiments at RHIC~\cite{STAR-1,STAR-2,STAR-3,STAR-4,STAR-5,STAR-6,STAR-7,STAR-8,Isobar-1,Isobar-2,STAR-27-CME-1} and the LHC~\cite{ALICE-1,CMS-1,CMS-2,ALICE-2,ALICE-3} without reaching a firm conclusion.
The occurrence of the CME requires the simultaneous fulfillment of several preconditions, which relies heavily on a magnetic field present at the time of QGP formation.
It is even possible that the residual strength of the magnetic field after pre-equilibrium evolution could still be strong at RHIC, but negligible at the LHC~\cite{DynamicalEvolution}.
Recent STAR data unveiled the magnetic field's imprint on the QGP through the deflection of charged particles in Au+Au collisions, with the effect being much stronger at a center-of-mass energy ($\sqrt{s_{NN}}$) of 27 GeV than at 200 GeV~\cite{STAR-Bfield}. 
This warrants systematically analyzing the RHIC Beam Energy Scan (BES) data, which could offer an enhanced opportunity to detect the CME~\cite{STAR-BESI-mini-review}. 
In particular, the BES-II datasets benefit from having over ten times higher statistics than BES-I and a dedicated detector (the event plane detector) for reconstructing the \(\vec{B}\) direction. 
Moreover, the halt of deconfinement and/or chiral symmetry restoration at sufficiently low $\sqrt{s_{NN}}$ should provide a pure-background scenario to test a CME observable.  

On average, the magnetic field is perpendicular to the reaction plane (RP), spanned by the impact parameter and the colliding beam direction.
To quantify the collective motion, incorporating the CME-induced charge separation, we apply a Fourier expansion to the azimuthal angle ($\varphi$) distribution of final-state charged particles relative to the RP ($\Psi_{\rm RP}$)~\cite{Sergeiflow1, CME-8}, 
\begin{equation}
\frac{dN_{\pm}}{d\varphi} \propto 1+ 2a_{1}^{\pm}\sin\Delta\varphi + \sum_{n=1}^\infty 2v_{n}^{\pm}\cos n\Delta\varphi , 
\label{equ:Fourier_expansion}
\end{equation}
where $\Delta\varphi = \varphi-\Psi_{\rm{RP}}$. $a_1^{\pm}$ and $v_n^{\pm}$ are Fourier coefficients corresponding to different collective motions for positively and negatively charged particles.  
The $v_2$ coefficient, known as elliptic flow, is believed to develop through
hydrodynamic expansion from the initial overlap geometry (eccentricity)~\cite{hydro} of the heavy-ion collision zone, and manifests itself as an ellipsoidal emission pattern of final-state particles.
The coefficients $a_{1}^{+}$ and $a_{1}^{-}$ are expected to have 
opposite signs due to the CME-induced charge separation~\cite{CME-8}. 
Since $\mu_5$ fluctuates event by event, the average value of $a_1^\pm$ will be be zero. 
Therefore, most CME observables aim to measure the fluctuation of $a_1$.  
A recent study~\cite{CME-9} establishes that the key components are equivalent between different charge separation observables, including the $\gamma^{112}$ correlator~\cite{CME-8}, the $R_{\Psi_2}(\Delta S)$ correlator~\cite{Magdy}, and the signed balance functions~\cite{Tang}. 
In this work, we employ  $\gamma^{112}$~\cite{CME-8} to measure the charge separation,
\begin{equation}
\gamma^{112} \equiv \langle \cos (\varphi_\alpha + \varphi_\beta - 2\Psi_{\rm{RP}}) \rangle,   
\label{eq:gamma112}
\end{equation}
where $\alpha$ and $\beta$ represent charge sign. 
The average is taken over particle pairs and then over events. 
We expect the CME effect to lead to a positive difference between opposite-sign (OS) and same-sign (SS) pairs
\begin{equation}
    \Delta\gamma^{112} \equiv \gamma_{\rm{OS}}^{112} - \gamma_{\rm{SS}}^{112}.
\end{equation}
Taking the difference eliminates the charge-independent background related to momentum conservation~\cite{STAR-3,STAR-4}. 
Positive $\Delta\gamma^{112}$ values have been reported by experiments at RHIC~\cite{STAR-1,STAR-2,STAR-3} and the LHC~\cite{ALICE-1}. 
However, data interpretation is complicated by charge-dependent backgrounds arising from elliptic flow coupled with mechanisms that correlate two particles, such as resonance decays~\cite{CME-8}, local charge conservation (LCC)~\cite{PrattSorren:2011,LCC-1}, and transverse momentum conservation (TMC)~\cite{Pratt2010,Flow_CME}.  
Unrelated to the RP, nonflow correlations~\cite{STAR-v2-first} can also supply a positive contribution to $\Delta\gamma^{112}$ from various sources including clusters, resonances, jets, and dijets. 

While nonflow effects can be minimized by reconstructing the event plane (an estimate of the RP) with the spectator particles, more effort is demanded to mitigate flow-related backgrounds.
One strategy is to compare two analogous measurements with different CME fractions ($f_{\rm CME}$) in $\Delta\gamma^{112}$. 
The RHIC isobar-collision program contrasts  $^{96}_{44}$Ru+$^{96}_{44}$Ru and $^{96}_{40}$Zr+$^{96}_{40}$Zr, which are presumed to exhibit distinct CME signals with similar backgrounds~\cite{isobar_proposal1,isobar_proposal2,isobar_proposal3}. However, the isobar data manifest no expected
CME signatures at $\sqrt{s_{NN}}$ = 200 GeV~\cite{Isobar-1,Isobar-2}.
At LHC energies, the CMS collaboration juxtaposes the $\Delta\gamma^{112}$ results in Pb+Pb and $p$+Pb, with the latter believed to be insensitive to the CME~\cite{CMS-1, CMS-2}.
Similar results observed in the two systems at the same multiplicity indicate the background dominates in Pb+Pb collisions at $\sqrt{s_{NN}} = 5.02$ TeV. 
Additionally, one can measure $f_{\rm CME}$ by comparing the observables with respect to two types of event planes, the participant and spectator planes~\cite{TwoPlane1,TwoPlane2}, or in two separate regions of pair invariant mass~\cite{InvMass}. 
The $f_{\rm CME}$ values measured by the STAR collaboration are consistent with zero in Au+Au at 200 and 27 GeV~\cite{STAR-8,STAR-7,STAR-27-CME-1}.

Instead of capitalizing on subtle distinctions between two similar measurements, an alternative strategy is to directly subtract the flow-related background. 
In reporting the BES-I data in 2014, STAR introduced a background subtraction scheme~\cite{STAR-4}
\begin{eqnarray}
\gamma^{112} &\equiv& \langle \cos (\varphi_\alpha + \varphi_\beta - 2\Psi_{\rm{RP}}) \rangle = \kappa_{{\rm{bg}}}^{112} v_2 F - H^{112}, \\
\label{eq:3}
\delta &\equiv& \langle \cos(\phi_\alpha -\phi_\beta) \rangle = F + H^{112},
\label{eq:4}    
\end{eqnarray}
where $H^{112}$ and $F$ are the CME and background contributions, respectively, and $\kappa_{\rm{bg}}^{112}$ is an adjustable parameter. $\delta$ represents the two-particle correlation. After canceling out $F$, we obtain

\begin{equation}
H^{112}(\kappa_{\rm{bg}}^{112}) \equiv (\kappa_{\rm{bg}}^{112} v_2 \delta - \gamma^{112})/(1+\kappa^{112}_{\rm{bg}} v_2).   
\label{h-corr}
\end{equation}

With an assumption of $\kappa_{\rm{bg}} \approx$ 2--3, the difference $\Delta H^{112} \equiv H^{112}_{\rm SS}-H^{112}_{\rm OS}$ tends to vanish in the 0--60\% centrality region of Au+Au collisions at $\sqrt{s_{NN}} = 7.7$ and 200 GeV and remains positive finite at intermediate beam energies~\cite{STAR-4}. 
This rise-and-fall trend is enticing, but the $\kappa_{\rm{bg}}^{112}$ value is not predetermined.

Another approach, event shape engineering (ESE)~\cite{SergeiESE}, classifies events based on the particle emission pattern and projects $\Delta\gamma^{112}$ onto isotropic eccentricity by selecting events with zero $v_2$.
Recent studies~\cite{rho00_dg112,rho00_dg112_cont} suggest that globally spin-aligned vector mesons could influence the $\Delta\gamma^{112}$ measurement through their decay daughters. 
As both $v_2$~\cite{rho00_v2} and $\Delta\gamma^{112}$ change monotonically with spin alignment, the background in $\Delta\gamma^{112}$  due to spin alignment should also vanish at zero $v_2$.
The ESE results reveal no CME signals at top RHIC energy~\cite{STAR-5,Xu:2023wcy} or LHC energies~\cite{ALICE-2,ALICE-3,CMS-2}. 
Notably, the construction of these event shape variables focuses on controlling eccentricity and excludes particles of interest (POI). 
This leads to a long extrapolation of $\Delta\gamma^{112}$ across a broad, unmeasured $v_{2,{\rm POI}}$ region, which introduces substantial statistical and systematic uncertainties~\cite{ESS-2}.

In this work, we apply a novel event shape selection (ESS) method ~\cite{1st-ESS,ESS-1,ESS-2} to the RHIC data. 
Finite initial-state eccentricity causes an ellipsoidal emission distribution and creates a long-range correlation in the flow planes along the beam direction.
However, even with identical eccentricity, the emission pattern can fluctuate considerably from event to event and across different rapidities within the same event~\cite{berndt-2}.
This short-range fluctuation in the emission pattern suggests that the event shape variable from a kinematic region other than the POI is not particularly effective in reflecting the magnitude of the flow-related background.
An event shape variable based on POI captures both initial geometry and final-state emission~\cite{ESS-2} and
provides better leverage to map onto isotropic events. 
To avoid a self-correlation with $v_{2,{\rm POI}}$, we construct the event shape variable using particle pairs of interest (PPOI) instead of individual particles~\cite{ESS-2}.

We also exploit another correlator~\cite{ESS-1,Subikash}
\begin{equation}
\gamma^{132} \equiv \langle \cos (\varphi_\alpha - 3\varphi_\beta + 2\Psi_{\rm{RP}}) \rangle   
\label{eq:gamma132}
\end{equation}
as a background indicator to validate the reduction of flow-related backgrounds.
The difference $\Delta\gamma^{132} \equiv \gamma_{\rm{OS}}^{132} - \gamma_{\rm{SS}}^{132}$ is dictated by background mechanisms akin to those for $\Delta\gamma^{112}$ and contains a negligible  CME contribution.
Analytical derivations and model calculations suggest the relation $\Delta\gamma^{132} = v_2 \Delta\delta$~\cite{Subikash}, where $\Delta\delta\equiv \delta_{\rm OS}-\delta_{\rm SS}$.
Furthermore, we can replace $H^{112}$ in Eq.~(\ref{h-corr}) with the ESS result for  $\Delta\gamma^{112}$ and solve for $\kappa^{112}_{\rm{bg}}$.
This enables us to understand the background mechanism in $\Delta\gamma^{112}$ in terms of a coupling between elliptic flow $v_2$ and two-particle correlations $\Delta\delta$. Similar to Eq.~(\ref{h-corr}), we can then define $H^{132}$ and derive the corresponding $\kappa_{\rm{bg}}^{132}$ for $\gamma^{132}$.

This paper is an expanded version of Ref.~\cite{short-letter} and is organized as follows.
Section~\ref{detector} describes the STAR detector, data collection, and our selection criteria for events and POI. Section~\ref{sec:anaMth} outlines the analysis method and systematic uncertainties.
Section~\ref{measurements-star} presents the STAR data of Au+Au collisions at eight beam energies from 7.7 to 200 GeV for $v_2$, $\Delta\gamma^{112}$, and $\Delta\gamma^{132}$, as well as the corresponding ESS results. 
Based on these results, we also estimate $f_{\rm CME}$ and extract $\kappa_{\rm{bg}}^{112}$ and $\kappa_{\rm{bg}}^{132}$. 
We draw brief conclusions in Section~\ref{result}.

\section{EXPERIMENT AND DATA SELECTION}\label{detector}

The STAR complex comprises a series of detector subsystems located in both the midrapidity and forward/backward-rapidity regions. 
This work mainly involves the Time Projection Chamber (TPC)~\cite{TPC-1}, the Time-of-Flight (TOF) system~\cite{tof}, the Vertex Position Detector (VPD)~\cite{VPD}, the Event Plane Detector (EPD)~\cite{EPD-1}, the Zero Degree Calorimeter (ZDC)~\cite{ZDC}, and the Beam-Beam Counter (BBC)~\cite{BBC}. 
Encased in a solenoidal magnet that provides a uniform magnetic field along the beam direction, the TPC tracks charged-particle helices at midrapidities with complete azimuthal coverage. 
The curvature information in the magnetic field yields the particle's momentum and charge sign.
The mean ionization energy loss per unit track length $\langle dE/dx \rangle$ is used to identify the particle species.  For a specific particle type $i$, we define the difference, in
terms of standard deviations between the measured $\langle dE/dx \rangle$ and the theoretical expectation from Bichsel function $\langle dE/dx \rangle_i^{\rm th}$~\cite{nSigma1,nSigma2}, 
\begin{equation}
n\sigma_i = \frac{1}{\sigma_R} \ln\bigg(\frac{\left< dE/dx \right>}{\left< dE/dx \right>^{\rm th}_i}\bigg).
\label{eq:nsigma}
\end{equation} 
Here, $\sigma_R$ is the momentum-dependent $\langle dE/dx \rangle$ resolution.

This analysis uses Au+Au collisions with a minimum-bias trigger that requires the coincidence of signals on both sides of the interaction region from the ZDC, the VPD, or the BBC. The RHIC BES-II data with $\sqrt{s_{NN}}$ from 7.7 to 27 GeV were collected between 2018 and 2021, and the 200 GeV data in 2016.
The total numbers of events analyzed for $\sqrt{s_{NN}} = 7.7$--27 (BES-II) and at 200 GeV are listed in Table~\ref{tab:epdcut}. 
As the TOF is a fast detector, we excluded out-of-time pile-up events by rejecting outlier events that show discrepancies between multiplicities counted by the TPC and the TOF. 
The primary vertex position for each event was determined by using the TPC data to find the vertex location along the beam direction ($V_{z,{\rm TPC}}$) and perpendicular to the beam line ($V_r$). 
We selected events with $|V_{z,{\rm TPC}}| < 70$ cm and $|V_r| < 2 $ cm. 
At 200 GeV, the $V_z$ cut was tightened to $|V_{z,{\rm TPC}}| < 30$ cm and $|V_{z,{\rm TPC}}-V_{z,{\rm VPD}}| <$ 3 cm, where $V_{z,{\rm VPD}}$ was determined by the VPD to reduce the beam-induced background.
Events are categorized into different centrality classes---0\%--5\%, 5\%--10\%, 10\%--20\%, 20\%--30\%, 30\%--40\%, 40\%--50\%, 50\%--60\%, 60\%--70\%, and 70\%--80\%---based on the uncorrected charged-particle multiplicity distribution at midrapidity, as detailed in Ref.~\cite{STAR-11}.

\begin{table}[ht]
    \caption{Total number of events $N_{E}$, beam rapidity, and the EPD or ZDC-SMD $\eta$ coverage for the spectator plane at each $\sqrt{s_{NN}}$ in BES-II and at 200 GeV.}
    \centering
    \begin{tabular}{ |c|c|c |c|c|c|c |c| c|}
\hline
\hline
$\sqrt{s_{NN}}$ (GeV) & 7.7 &9.2 & 11.5 & 14.6 &17.3 & 19.6 & 27 &200 \\
\hline
$N_{E}$ ($\times10^6$)& 60  & 110 & 160 & 230 & 200 & 360 & 490 & 2000 \\
\hline
$y_{\rm beam}$ & 2.09 &2.27 & 2.50 & 2.74 & 2.91 & 3.04 & 3.36  & 5.36\\
\hline
$|\eta| >$ & 2.3 &2.5 & 2.7 & 3.0 & 3.1 & 3.2 & 3.8 & 6.3\\
\hline
\hline
    \end{tabular}
    \label{tab:epdcut}
\end{table}

We selected charged mesons using kinematic cuts on pseudorapidity, momentum, and transverse momentum of $|\eta| < 1$, $p < 1.4$ GeV/$c$, and $p_T > 0.2$ GeV/$c$, respectively. 
These kinematic cuts ensure that the dynamics are governed by hydrodynamics, as established through years of RHIC experimental measurements~\cite{BRAHMS-whitepaper,PHOBOS-whitepaper,STAR-whitepaper,PHENIX-whitepaper}.
To ensure track quality, we required at least 15 ionization points ($N_{\rm hits}$) be used in the helix reconstruction.
The distance of closest approach (DCA) to the primary vertex was restricted to less than 3 cm to reduce secondary particles from weak decays and contamination from interactions between primary particles and detector materials.
We preferentially selected mesons, mostly $\pi^\pm$ and $K^\pm$, and rejected (anti)protons from the analysis for several reasons. First, an analysis focusing solely on mesons is preferred over a mixture of particle types, allowing for a clear interpretation of the data and model comparison.
For example, CME expectations differ for baryons and mesons, and (anti)protons could be less than ideal for CME searches in high-energy heavy-ion collisions. 
Reference~\cite{CME-5} argues that in the two-flavor scenario, the CME-induced baryon number difference vanishes, whereas in the three-flavor case, the charge difference still primarily manifests in mesons. 
Baryons also exhibit distinct elliptic flow behavior compared with mesons~\cite{STAR-BESv2}, which complicates efforts to suppress $v_2$-related background effects when all charged hadrons are grouped. Furthermore, the $\bar{\rm p}/{\rm p}$ ratio is only around 0.01--0.1 at lower beam energies~\cite{par-p-ratio}, introducing notable asymmetry, whereas charge separation measurements ideally require equal positive and negative yields. 
Protons are more heavily influenced by transported quarks~\cite{STAR-BESv1} than mesons, affecting both the signal and the background. 
Finally, antiprotons are more easily absorbed in the baryon-rich environment than protons, adding further complications to both signal and background interpretation.

\begin{figure}[!tbhp]
\includegraphics[width=0.23\textwidth]{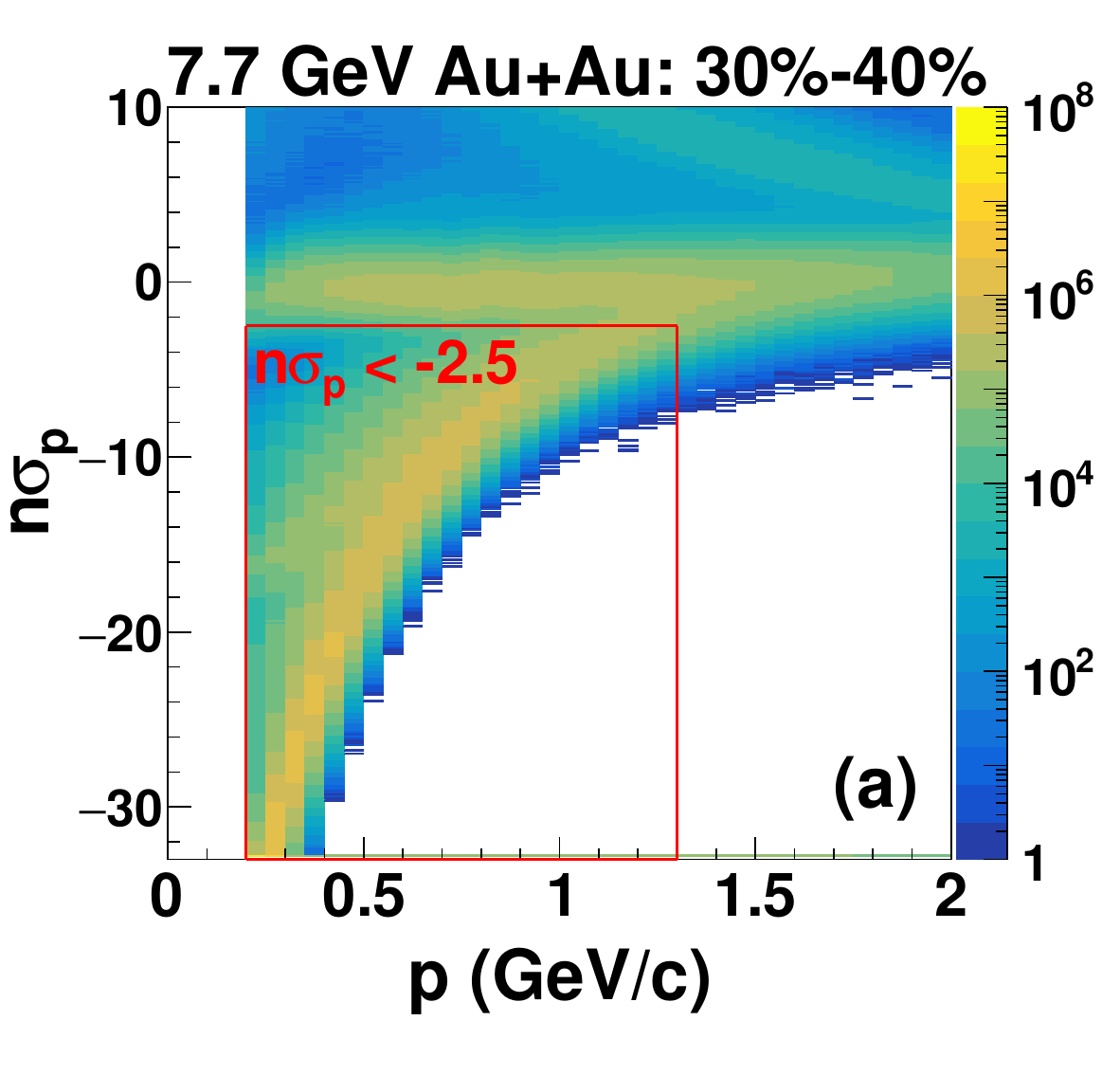}
\includegraphics[width=0.23\textwidth]{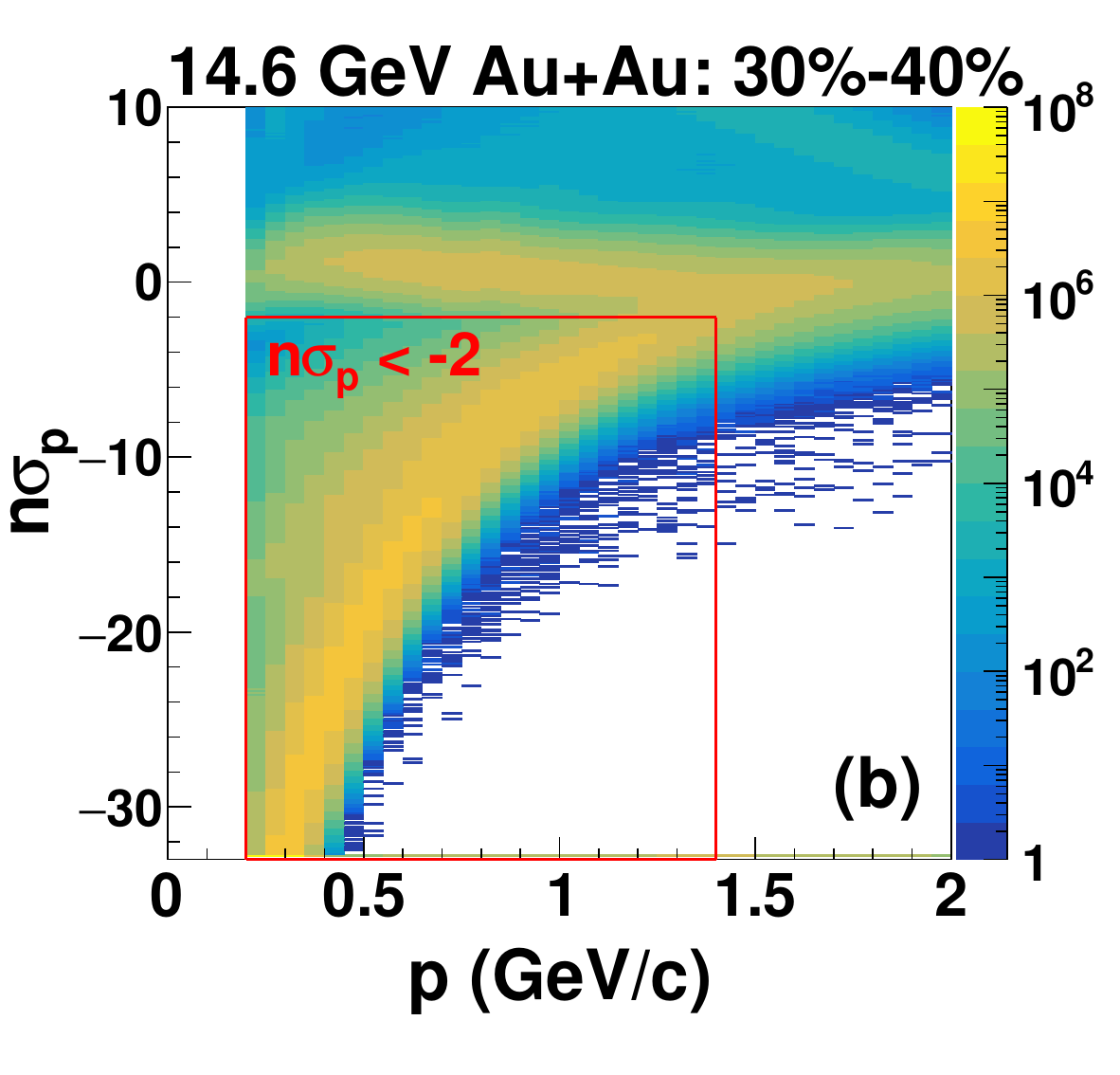}
\caption{ $n\sigma_{\rm{p}}$ (proton $n\sigma$ as defined in Eq.~(\ref{eq:nsigma})) vs momentum ($p$) for positively charged particles in the 30\%--40\% centrality region of Au+Au collisions at $\sqrt{s_{NN}} =$ 7.7 GeV (a) and 14.6 GeV (b).
The lower-left corners delineated with vertical and horizontal lines represent the selection of POI with $p < 1.3(1.4)$ GeV/$c$ and $n\sigma_{\rm{p}} < -2.5(2)$ to exclude (anti)protons.}
\label{nsigma}
\end{figure}

Figure~\ref{nsigma} illustrates the 2D distributions of $n\sigma_{\rm{p}}$ vs momentum for positively charged particles in 30\%--40\% Au+Au collisions at 7.7 GeV (a) and 14.6 GeV (b).
Above 7.7 GeV, we applied a cut of $n\sigma_{\rm{p}}< -2$ to exclude (anti)protons from the POI.
At 7.7 GeV, where protons are more abundant, we tightened the cut to $n\sigma_{\rm{p}}< -2.5$ and set the momentum cut to $p < 1.3$ GeV/$c$ accordingly.
The corresponding purity was higher than 99.9\%.
The data analyses were also corrected for the inefficiency due to the TPC tracking capabilities and the $n\sigma_{\rm{p}}$ cut. 
We evaluated the uncertainty associated with the efficiency correction and found its effect to be negligible compared with other sources of uncertainty.

\section{Analysis method} \label{sec:anaMth}
\subsection{Event planes}
Compared with the participant plane, the spectator plane is not only less sensitive to nonflow effects but should also be more closely associated with the $\vec{B}$ direction~\cite{TwoPlane1,TwoPlane2}. 
At 200 GeV, we reconstructed the forward and backward $1^{\rm st}$-order event planes ($\Psi_1^f$ and $\Psi_1^b$) using the spectator neutron information registered by the ZDC Shower-Maximum Detectors (ZDC-SMD, $|\eta| > 6.3$)~\cite{{v1_62GeV}}.
At lower beam energies, the ZDC-SMDs have low efficiency, so we used the EPD ($2.1<|\eta|<5.1$)~\cite{EPD-1} instead. 
To target the region rich in spectator protons we chose inner EPD tiles beyond beam rapidity ($|\eta| > y_\mathrm{beam}$), as listed in Table.~\ref{tab:epdcut}.
After raising the lower boundary of the EPD tiles by 0.2 units in $|\eta|$, the results remain highly consistent with the default, indicating that the default approach already ensures a sufficiently large $\eta$ gap to render nonflow effects negligible.

In general, the $1^{\rm st}$-order event plane from a hit-based detector is taken as the azimuthal angle of a flow vector $\bigl(\sum_{i} w_i x_i ,~\sum_{i} w_i y_i\bigr)$~\cite{Sergeiflow1}, where the weight $w_i$ accounts for the energy deposition in the location of a ZDC-SMD or EPD channel ($x_i,y_i$). 
When using an EPD tile, the directed flow information at the corresponding $\eta$ is also incorporated into $w_i$, following the procedures adopted in Ref.~\cite{STAR-27-CME-1}. 
The event planes thus obtained are further corrected to have uniform distributions according to the method described in Ref.~\cite{E877:1997zjw}. 

\begin{figure}[!tbhp]
\centering
\includegraphics[width=0.48\textwidth]{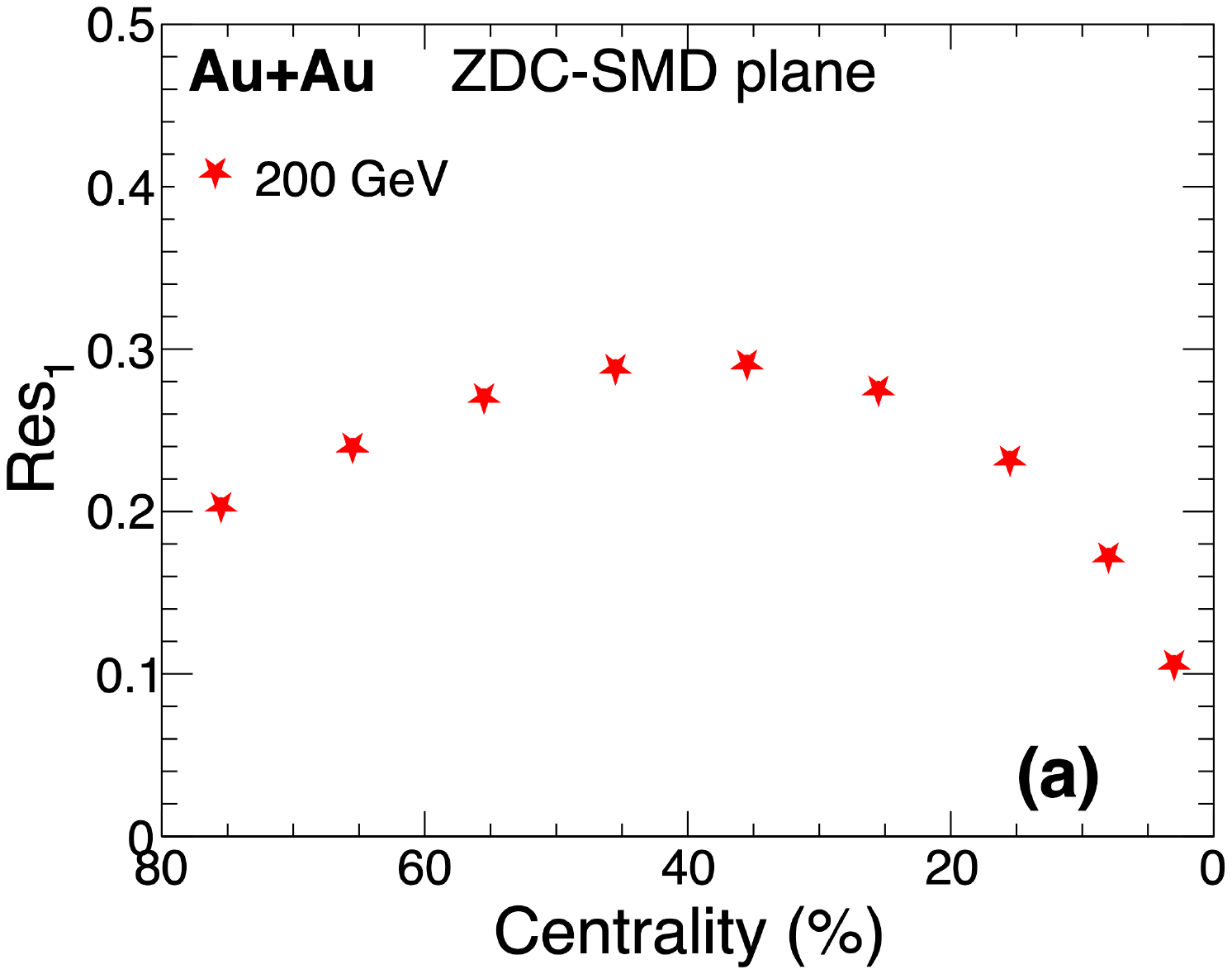}
\includegraphics[width=0.48\textwidth]{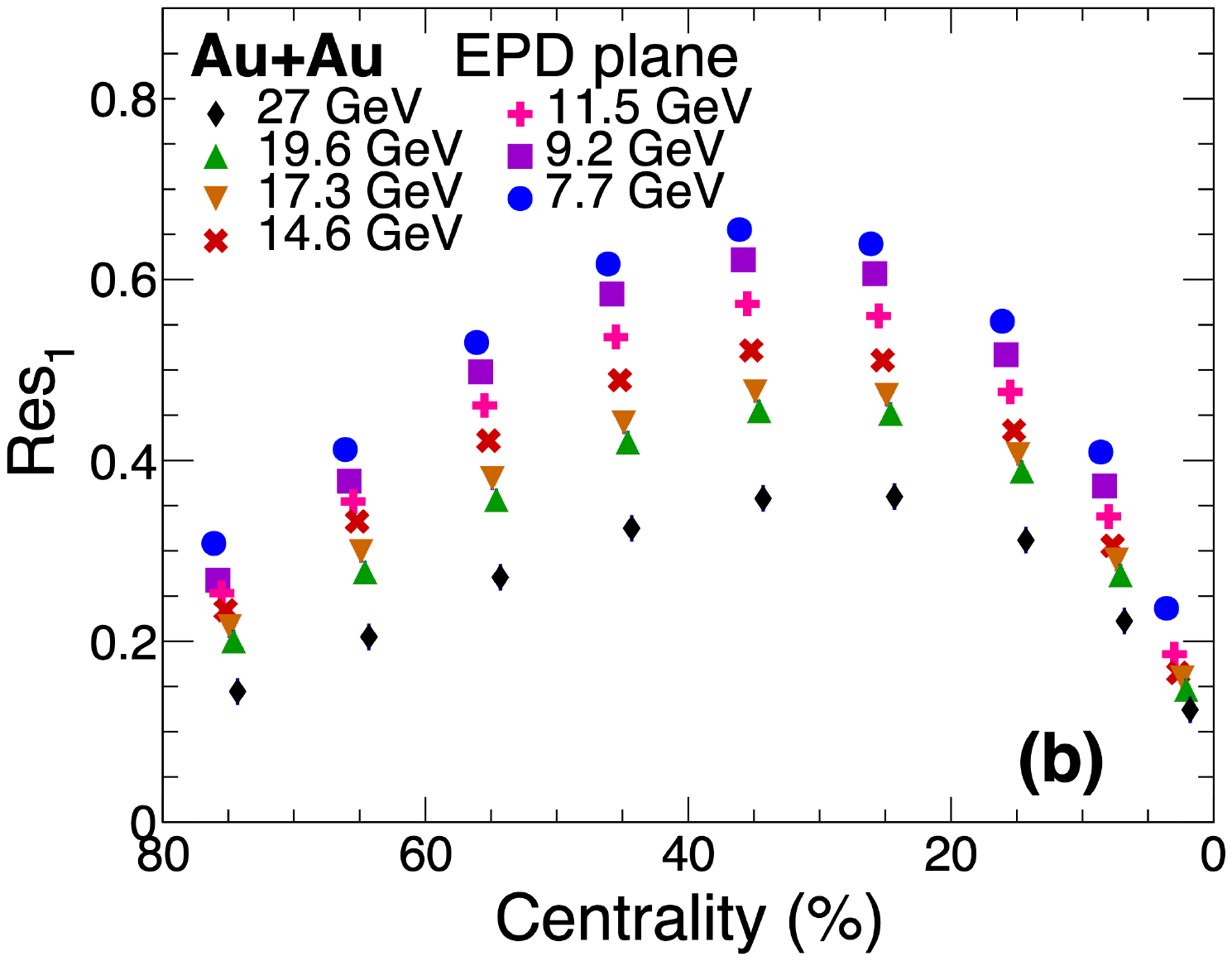}
\caption{Resolution of $1^{\rm st}$-order event planes ($\Psi_1^f$ or $\Psi_1^b$) as a function of centrality in Au+Au collisions.
The event planes were reconstructed from the ZDC-SMD at $\sqrt{s_{NN}}$ = 200 GeV (a), while the EPD was used at 7.7--27 GeV (b).
}
\label{EPres-EPD}
\end{figure}

Figure~\ref{EPres-EPD} shows the resolution of the $1^{\rm st}$-order event planes ($\Psi_1^f$ or $\Psi_1^b$)~\cite{Sergeiflow1} 
\begin{equation}
    Res_{1} = \sqrt{ \langle \cos(\Psi_{1}^{f} - \Psi_{1}^{b} )\rangle}, 
    \label{sub-event-resolution}
\end{equation}
as a function of centrality in Au+Au collisions. 
The resolution is influenced by both the magnitude of $v_1$ and the yield of particles involved in the reconstruction of the event plane, with their convolution resulting in dependence on centrality and beam energy. 

\subsection{Event shape selection}

\begin{figure*}[!tbhp]
\centering
\includegraphics[width=\textwidth]{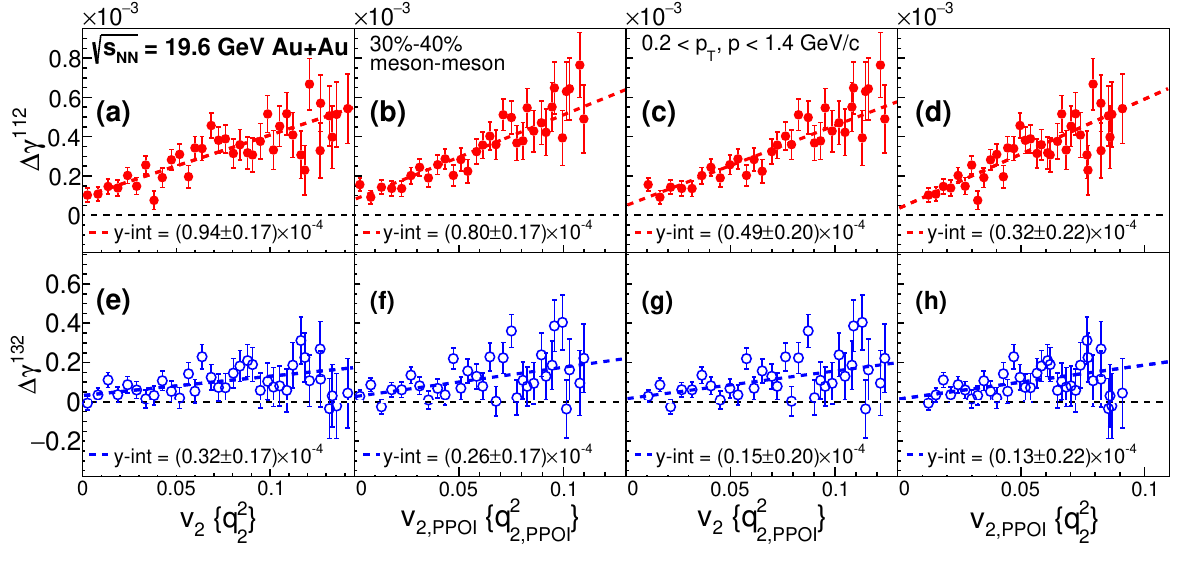}
\caption{The four ESS approaches that extrapolate $\Delta\gamma^{112}$ (a--d) and $\Delta\gamma^{132}$ (e--h) to zero $v_2$ or zero $v_{2,{\rm PPOI}}$ with events classified with either
$q^2_2$ or $q^2_{2,{\rm PPOI}}$ in 30\%--40\% Au+Au collisions at $\sqrt{s_{NN}} =19.6$ GeV.
The distinction between the panels in each row is indicated by the title on the $x$-axis.
The linear fits (dashed lines) are used to extract the $y$-intercepts at zero flow. For clarity of presentation, we combined every four data points in (b), (c), (g), and (h), but the linear fits were obtained from the data without rebinning.
}
\label{4ESS}
\end{figure*}

\begin{table*}[!htbp]
\caption{$N_{\mathrm{part}}$ as a function of centrality in Au+Au collisions at various collision energies.}
  \centering
\begin{tabular}{|c|c|c|c|c|c|c|c|c|}
\hline
\hline
\diagbox[innerwidth = 5cm, height = 4ex]{Centrality(\%)}{$\sqrt{s_{NN}}$ (GeV)}&7.7&9.2&11.5&14.6&17.3&19.6&27& 200\\ \hline
70-80&14.5&13.6&13.7&13.8&14.1&14.6&14.4&14.4\\ 
\hline
60-70&26.3&26.0&26.0&26.1&25.9&26.9&26.3&27.4\\ 
\hline
50-60&45.0&45.4&44.8&45.3&44.7&45.7&45.1&47.8\\ 
\hline
40-50&71.7&72.8&72.1&72.2&72.3&73.1&72.9&76.9\\ 
\hline
30-40&109.0&109.8&109.3&109.6&109.7&110.2&110.5&116.7\\ \hline
20-30&158.4&158.8&158.1&159.8&159.5&159.8&160.3&169.0\\ \hline
10-20&224.3&224.2&223.4&225.8&226.4&225.3&226.9&237.3\\ \hline
5-10&288.5&288.0&286.8&289.5&290.0&289.4&291.2&302.2\\ \hline
0-5&336.2&336.2&337.0&339.4&338.7&340.2&339.8&349.6\\ 
\hline
\hline
\end{tabular}
\label{npart_bes2}
\end{table*}

The ESS method used here differs from the ESE primarily in the definition of the event shape variable. While the ESE constructs the event shape variable from a specific kinematic region by excluding the POI and relies on long-range flow correlations, this approach can be compromised by flow-plane decorrelations~\cite{rn-CMS, decor-ATLAS, decor-ATLAS2, Maowu}. Additionally, the ESE method involves significant extrapolation to zero $v_2$, which introduces substantial uncertainties~\cite{ESS-2}. In contrast, the ESS method~\cite{1st-ESS,ESS-1,ESS-2} 
constructs the event shape variables directly from POI or PPOI,
which captures both the initial geometry of collision and expansion properties of QGP.
This approach requires significantly less extrapolation to zero $v_2$ compared with the ESE method, resulting in much higher precision~\cite{ESS-2}.

With respect to the $1^{\rm st}$-order event planes $\Psi^f_1$ and $\Psi^b_1$, we mainly measured three observables 
\begin{eqnarray}
v_2 &=& \langle \cos(2\varphi-\Psi^f_1-\Psi^b_1)\rangle/\langle \cos(\Psi_{1}^{f} - \Psi_{1}^{b} )\rangle, \label{v2-eqt}\\ 
\gamma^{112} &=& \frac{\langle \cos(\varphi_\alpha+\varphi_\beta-\Psi^f_1-\Psi^b_1)\rangle}{\langle \cos(\Psi_{1}^{f} - \Psi_{1}^{b} )\rangle}, \\ 
\gamma^{132} &=& \frac{\langle \cos(\varphi_\alpha-3\varphi_\beta+\Psi^f_1+\Psi^b_1)\rangle}{\langle \cos(\Psi_{1}^{f} - \Psi_{1}^{b} )\rangle}. 
\end{eqnarray}
Here, we use a variation of the event-plane method that automatically accounts for the possibility of a slight difference in resolution between $\Psi_{1}^{f}$ and $\Psi_{1}^{b}$. 
In addition to inclusive results from the ensemble averages, we also applied the ESS
approach to suppress the flow background by categorizing collision events based on the magnitude squared of a $2^{\rm nd}$-order flow vector~\cite{ESS-2},
\begin{eqnarray}
    q^2_{2}&=& \frac{\bigl(\sum^{N}_{i=1} \sin2\varphi_i\bigr)^2 + \bigl(\sum^{N}_{i=1} \cos2\varphi_i\bigr)^2 }{N(1+N v_{2}^2)}, \\
q^2_{2,{\rm PPOI}}&=& \frac{\bigl(\sum^{N_{\rm pair}}_{i=1} \sin2\varphi_i^{\rm p}\bigr)^2 + \bigl(\sum^{N_{\rm pair}}_{i=1} \cos2\varphi_i^{\rm p}\bigr)^2 }{N_{\rm pair}(1+N_{\rm pair} v_{2, \rm PPOI}^2)}, ~~
\end{eqnarray}
where $N$ and $N_{\rm pair}$ denote the multiplicities of POI and PPOI, respectively, in each event.
$\varphi_i^{\rm p}$ represents the azimuthal angle of a particle pair which was obtained by summing the momenta of two particles. 
$v_{2, \rm PPOI}^2$ is defined similarly to Eq.~(\ref{v2-eqt}), with $\varphi$ replaced with $\varphi^{\rm p}$.

In practice, we first used an event shape variable ($q_{2}^2$ or $q_{2,{\rm PPOI}}^2$) to categorize events into different subsets ($j$). 
In each subset, $\Delta\gamma^{112}_j$, $\Delta\gamma^{132}_j$, and the elliptic flow variable ($v_{2,j}$ or $v_{2,{\rm PPOI},j}$) were measured independently. 
Next, we plotted $\Delta\gamma^{112}_j$ and $\Delta\gamma^{132}_j$ against $v_{2,j}$ (or $v_{2,{\rm PPOI},j}$) and derived the $y$-intercept at zero $v_2$ (or $v_{2,{\rm PPOI}}$) to remove the flow-related background.
Since both the event shape variable and the elliptic flow variable can be calculated from single particles or from particle pairs, 
there are four recipes for extracting the $y$-intercept of $\Delta\gamma^{112}$ or $\Delta\gamma^{132}$.

Figure~\ref{4ESS} demonstrates the application of the four ESS recipes to $\Delta\gamma^{112}$ and $\Delta\gamma^{132}$ in 30\%--40\% Au+Au collisions at $\sqrt{s_{NN}} =19.6$ GeV.
In each case, we extracted the $y$-intercept using a linear fit (dashed line). 
The projection takes advantage of data points with elliptic flow as close to zero as possible, ensuring the robustness of the extrapolation. 
The four $y$-intercepts exhibit a clear ordering, which agrees with the analytical predictions and corroborates the various model calculations in Ref.~\cite{ESS-2}.
In Figs.~\ref{4ESS}(a) and \ref{4ESS}(b), the event shape variable and the elliptic flow variable use the same particles or particle pairs, leading to small residual backgrounds from self-correlations~\cite{ESS-2}. 
Similarly, the finite $y$-intercepts in Figs.~\ref{4ESS}(e) and \ref{4ESS}(f) are attributed to this effect, whereas the ideal $y$-intercepts for the background indicator $\Delta\gamma^{132}_{\rm ESS}$ should be zero.
The residual background is suppressed in Figs.~\ref{4ESS}(c) and \ref{4ESS}(d) for $\Delta\gamma^{112}$, and in Figs.~\ref{4ESS}(g) and \ref{4ESS}(h) for $\Delta\gamma^{132}$, 
since the event shape variable and the elliptic flow variable are based upon different azimuthal angles:
one is derived from single particles and the other from particle pairs.
In the panels involving $v_{2,\rm PPOI}$, an over-subtraction of the background arises from the inherent correlation between $v_{2,\rm PPOI}$ and $\Delta\gamma^{112}$~\cite{ESS-2}.
In this following, we take the ESS recipe using $q^2_{2,\rm PPOI}$ and single-particle $v_2$ as the best solution.
Finally, we multiplied the $y$-intercept extracted from Fig.~\ref{4ESS} by a small correction $(1- \overline{v}_2)^2$ to recover the unbiased signal  $\Delta\gamma^{112}_{\rm{ESS}}$~\cite{ESS-2}, where $\overline{v}_2$ is an average over all subsets. The correction is attributed to non-interdependent collectivity and has been validated through model studies~\cite{ESS-2,non-inter-dept}. 

\begin{figure}[!tbhp]
\centering
\includegraphics[width=0.48\textwidth]{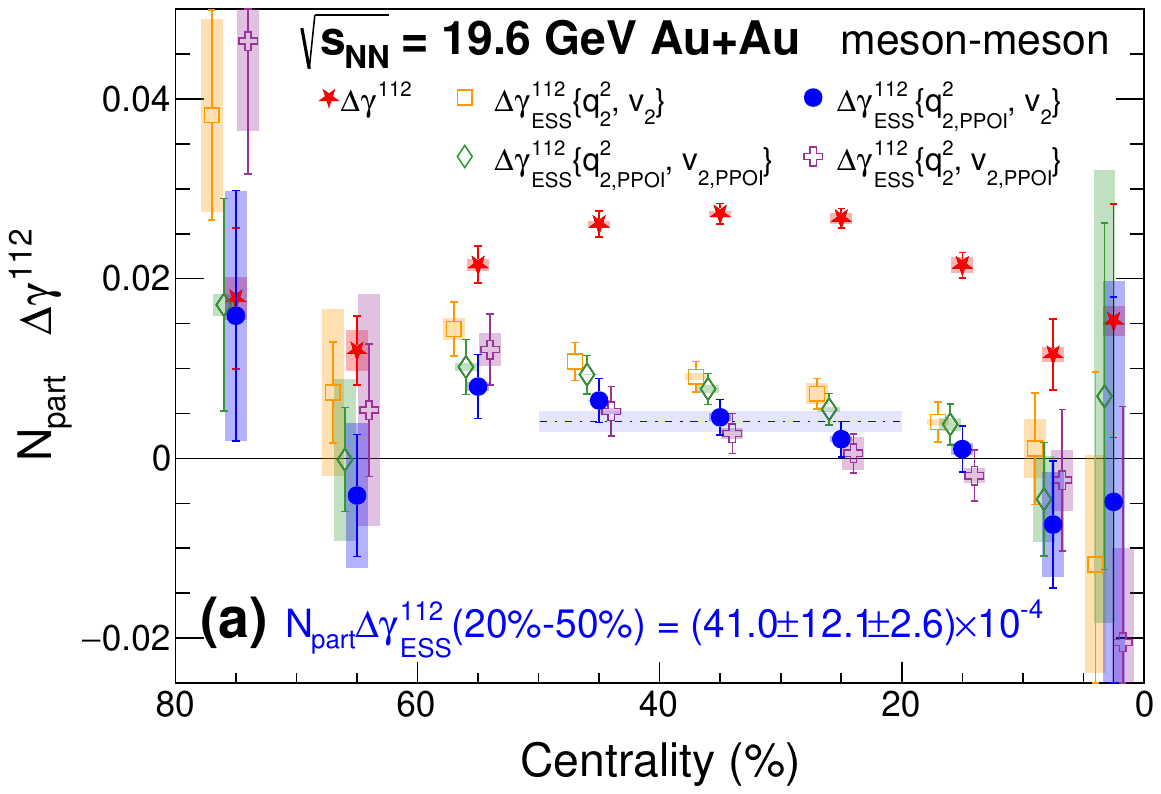}
\includegraphics[width=0.48\textwidth]{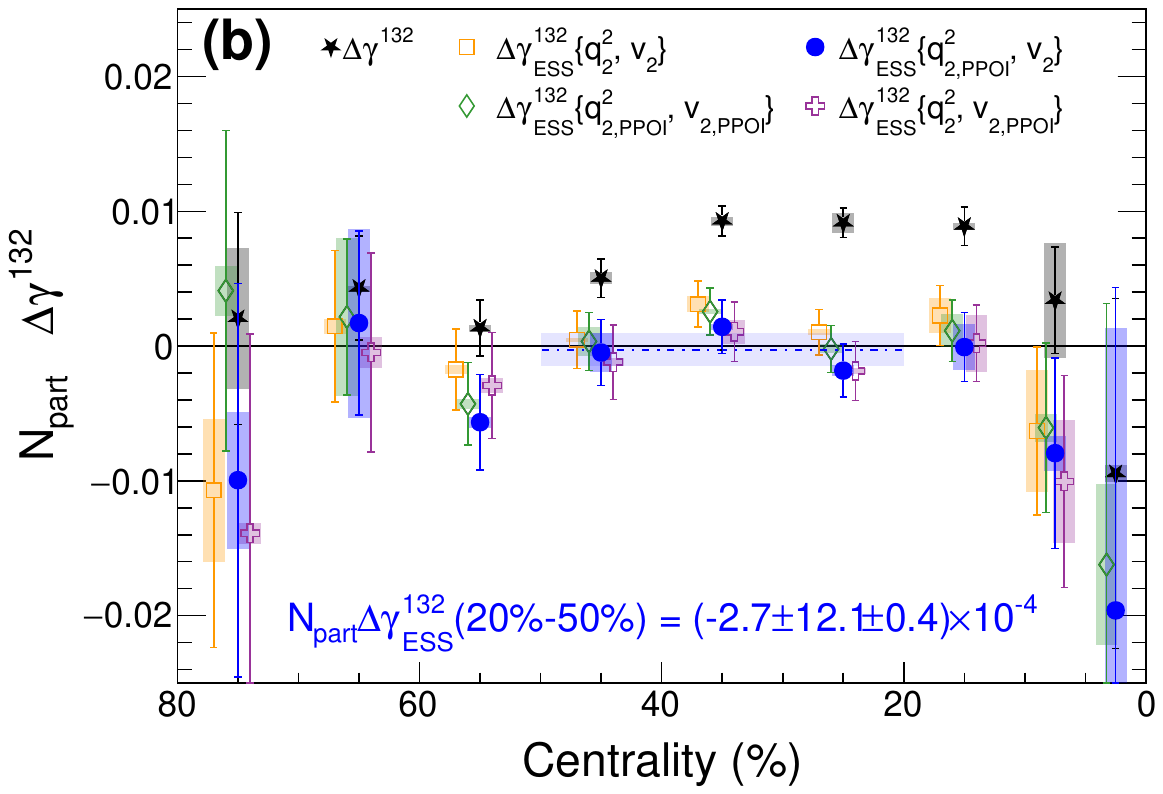}
\caption{Centrality dependence of (a) $N_{\rm part}\Delta\gamma^{112}$ and (b) $N_{\rm part}\Delta\gamma^{132}$ in comparison with the corresponding ESS results using four analysis recipes in Au+Au collisions at $\sqrt{s_{NN}} =19.6$ GeV. The vertical bars denote the statistical uncertainties, and the shaded boxes represent the systematic uncertainties. The horizontal dashed lines
denote constant fits to $N_{\rm part}\Delta\gamma^{112}_{\rm ESS}$ (or $N_{\rm part}\Delta\gamma^{132}_{\rm ESS}$) over the 20\%--50\% centrality region.}

\label{19-ESS}
\end{figure}

Figure~\ref{19-ESS} shows the data for Au+Au collisions at $\sqrt{s_{NN}} =19.6$ GeV as an example to illustrate the centrality dependence of (a) $N_{\rm part}\Delta\gamma^{112}$ and (b) $N_{\rm part}\Delta\gamma^{132}$.
The number of participating nucleons ($N_{\rm part}$), calculated using the Monte Carlo Glauber model (see Table~\ref{npart_bes2}), is used to compensate for the dilution effect on the correlation as a function of collision centrality~\cite{STAR-3}. 
For both observables, the ESS results exhibit a substantial reduction from the ensemble averages, with the ordering among the four recipes generally preserved. 
The same ordering is present at all the beam energies under study in this paper. 
(In the following, we will devote the subscript ``ESS" to the results obtained from the ESS recipe using $q^2_{2,\rm PPOI}$ and single-particle $v_2$).
In both the most central and most peripheral collisions, all of the results are consistent with zero within the statistical uncertainties.
In mid-central (20\%--50\%) collisions, the background indicator $N_{\rm part}\Delta\gamma^{132}_{\rm ESS} =[-2.7\pm12.1({\rm stat.})\pm0.4({\rm syst.})]\times 10^{-4}$ is consistent with zero, validating the background suppression achieved with this method.
In contrast, $N_{\rm part}\Delta\gamma^{112}_{\rm ESS} = [41.0\pm12.1({\rm stat.})\pm 2.6({\rm syst.})]\times 10^{-4}$ indicates a finite signal with a 3.3$\sigma$ significance in the 20\%--50\% centrality region.

\subsection{Systematic uncertainties}

Statistical uncertainties dominate over systematic ones in this analysis.
The procedure to assess the systematic uncertainties follows the methods outlined in Ref.~\cite{Isobar-1}.
Briefly, we adjust the following cuts to evaluate the uncertainty due to the TPC performance:
$0 < V_z < 70$ cm ($0 < V_z < 30$ cm for 200 GeV), $N_{\rm hits}\ge20$, DCA $< 1$ cm. 
We also included a track-splitting check by requiring the ratio of $N_{\rm hits}$ to the maximum number of ionization points to be greater than 0.52 for each track.
Additionally, the effect of (anti)proton rejection was investigated with a tighter cut of $n\sigma_{\rm{p}}<-3$ ($-3.5$ for 7.7 GeV). 

We denote $\Delta_i$ as the difference between the results obtained with the default and the varied cuts. 
For each variation $i$, we employed the Barlow procedure~\cite{Barlow-1} to evaluate the impact of the statistical uncertainties. Here, $\sigma_{{\rm stat},d}$ and $\sigma_{{\rm stat},i}$ represent the statistical uncertainties associated with results from the default and the varied cuts, respectively.
If the difference squared $\Delta_i^2$ is larger than $|\sigma^2_{{\rm stat},i} - \sigma^2_{{\rm stat},d}|$, we define $\sigma_i = \sqrt{\Delta_i^2 - |\sigma^2_{{\rm stat},i} - \sigma^2_{{\rm stat},d}|}$.
Otherwise, $\sigma_i = 0$. 
The minus sign in $|\sigma^2_{{\rm stat},i} - \sigma^2_{{\rm stat},d}|$ arises because the statistics for the varied cut are a subset of those for the default one. 
The overall systematic uncertainty is computed as the quadrature sum of $\sigma_i/\sqrt{12}$ (or $\sigma_i/\sqrt{3}$ for the $V_z$ variation), assuming the default and varied-cut results represent the maximum and minimum (or the central value and one extreme in the case of $V_z$), regardless of order, under a flat prior~\cite{FlatPrior} for each source. For each systematic source, we calculate the average of the two extreme values and report the mean of these five averages as our final central value. We performed a $\chi^2$ test~\cite{Lyons} at 19.6 and 200 GeV with additional DCA cut variations at $<1.5$, $<2$, and $<2.5$ cm, and found that the resulting systematic uncertainties are comparable to those obtained using the flat prior.
The primary sources of systematic uncertainties were the cuts on DCA and $n\sigma_{\rm{p}}$, each typically contributing at the level of 10\% relative to the corresponding statistical uncertainty.
The variations in the remaining cuts had marginal effects.

\begin{figure}[!tbhp]
\centering
\includegraphics[width=0.48\textwidth]{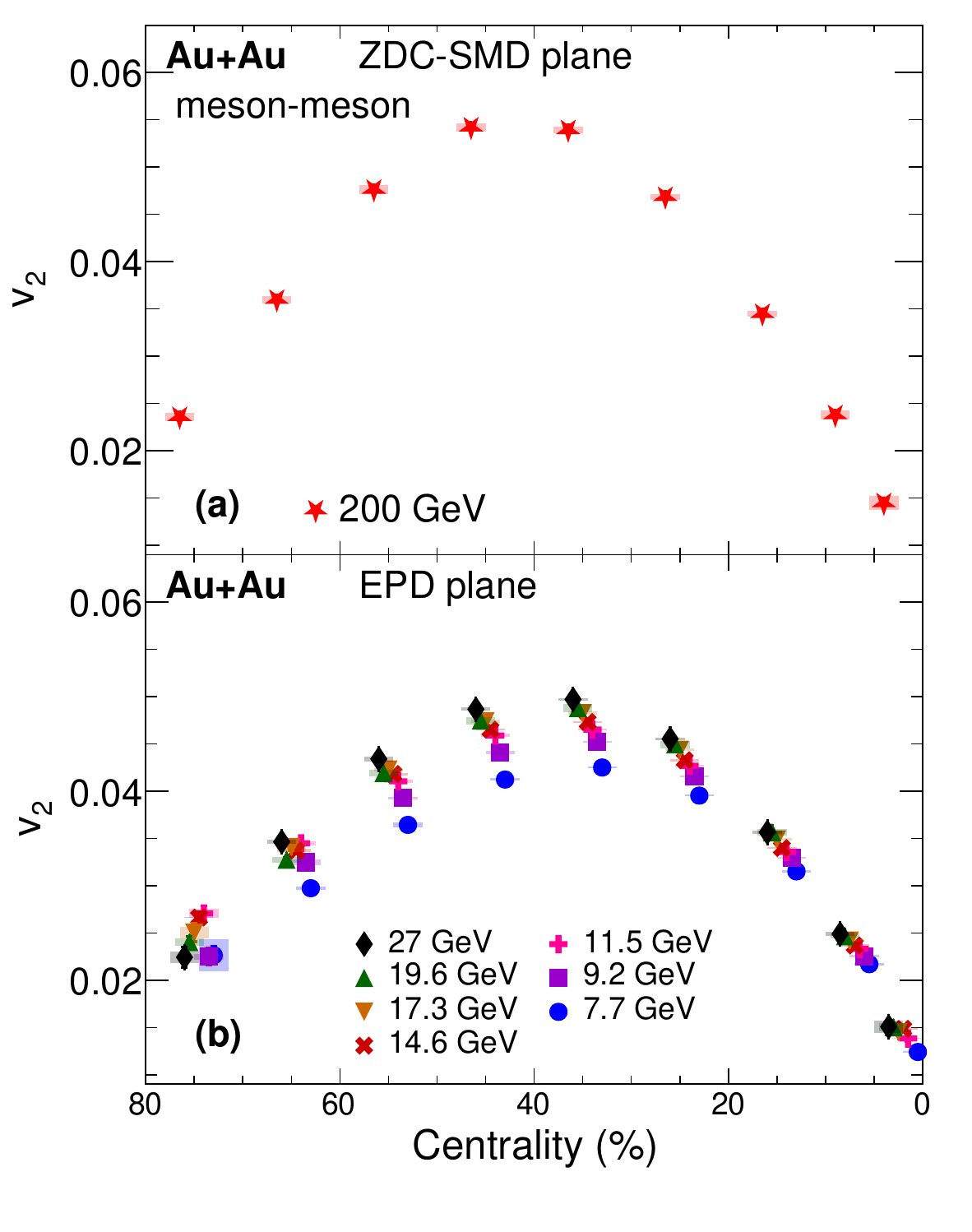}
\caption{$v_2$ of charged mesons as a function of centrality in Au+Au collisions measured with respect to the event plane from the ZDC-SMD at $\sqrt{s_{NN}}$ = 200 GeV (a) and the EPD at 7.7--27 GeV (b).}
\label{v2-EPD}
\end{figure}

\begin{figure*}[!tbhp]
\centering
\includegraphics[width=\textwidth]{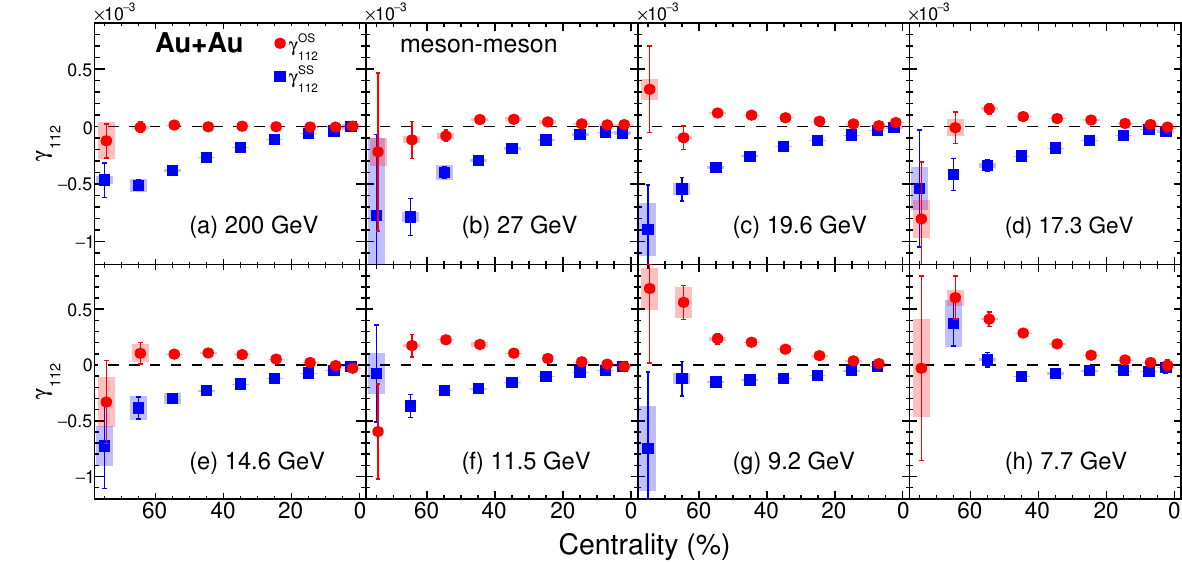}
\caption{Centrality dependence of 
$\gamma^{112}_{\rm OS}$ (red circles) and $\gamma^{112}_{\rm SS}$ (blue squares) in Au+Au collisions at $\sqrt{s_{NN}}$ = 7.7--200 GeV. 
The solid error bars and the shaded boxes represent the statistical and systematic uncertainties, respectively. 
}
\label{dg112}
\end{figure*}

\begin{figure*}[tbhp]
\centering
\includegraphics[width=\textwidth]{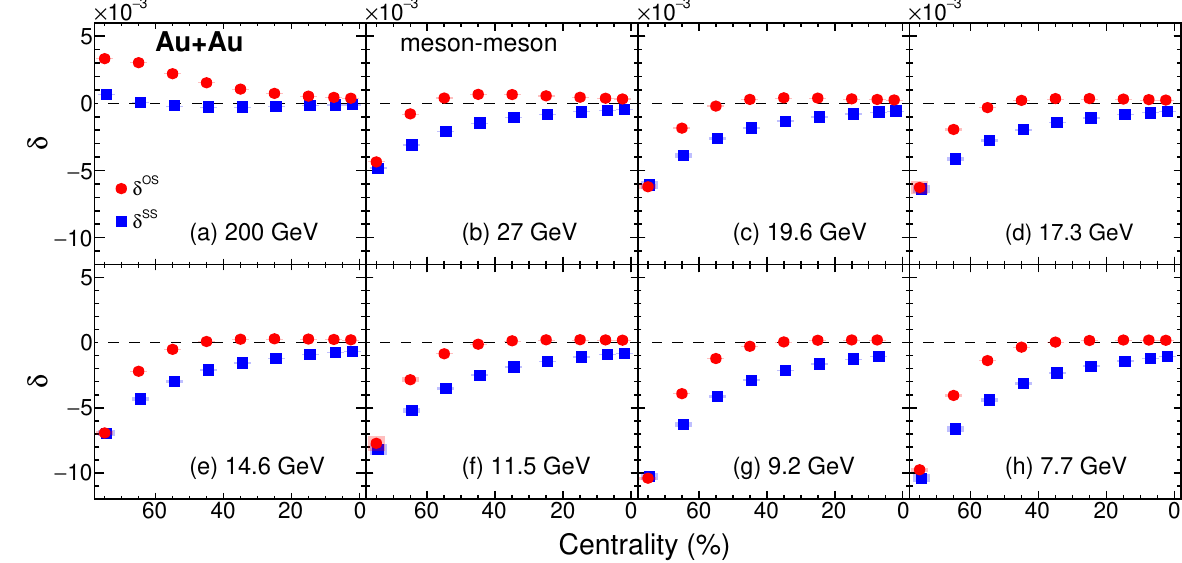}
\caption{Same as Fig.~\ref{dg112}, but for $\delta_{\rm OS}$ (red circles) and $\delta_{\rm SS}$ (blue squares).
}
\label{ddelta}
\end{figure*}

\begin{figure*}[tbhp]
\centering
\includegraphics[width=\textwidth]{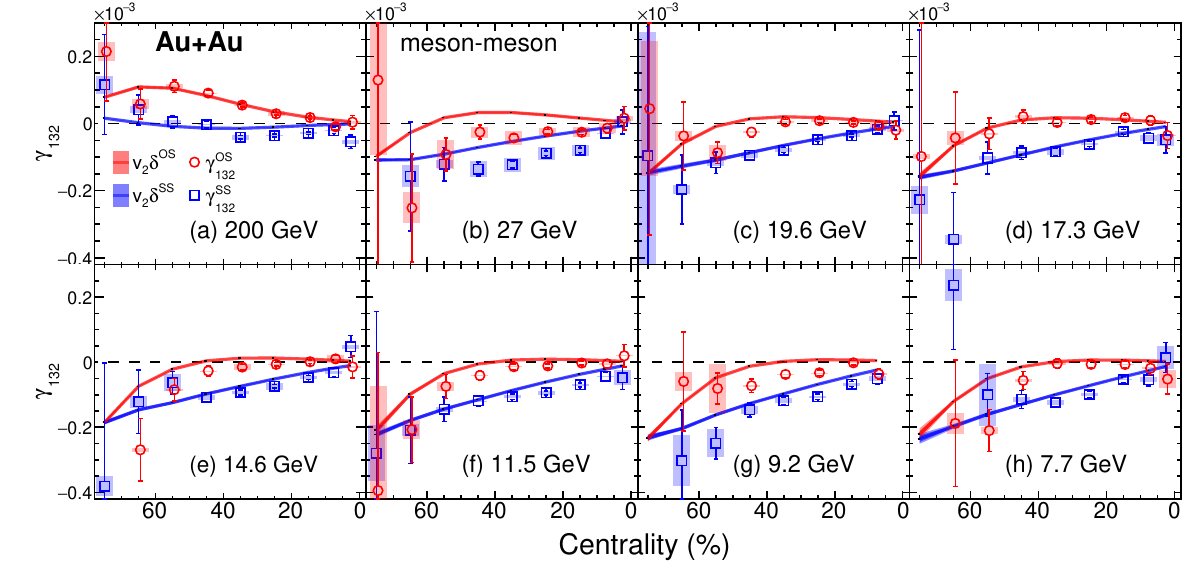}
\caption{Same as Fig.~\ref{dg112}, but for  
$\gamma^{132}_{\rm OS}$ and $\gamma^{132}_{\rm SS}$. For comparison, we also add $v_2\delta_{\rm OS}$ and $v_2\delta_{\rm SS}$ (shaded bands).
}
\label{dg132}
\end{figure*}

\begin{figure*}[tbhp]
\centering
\includegraphics[width=\textwidth]{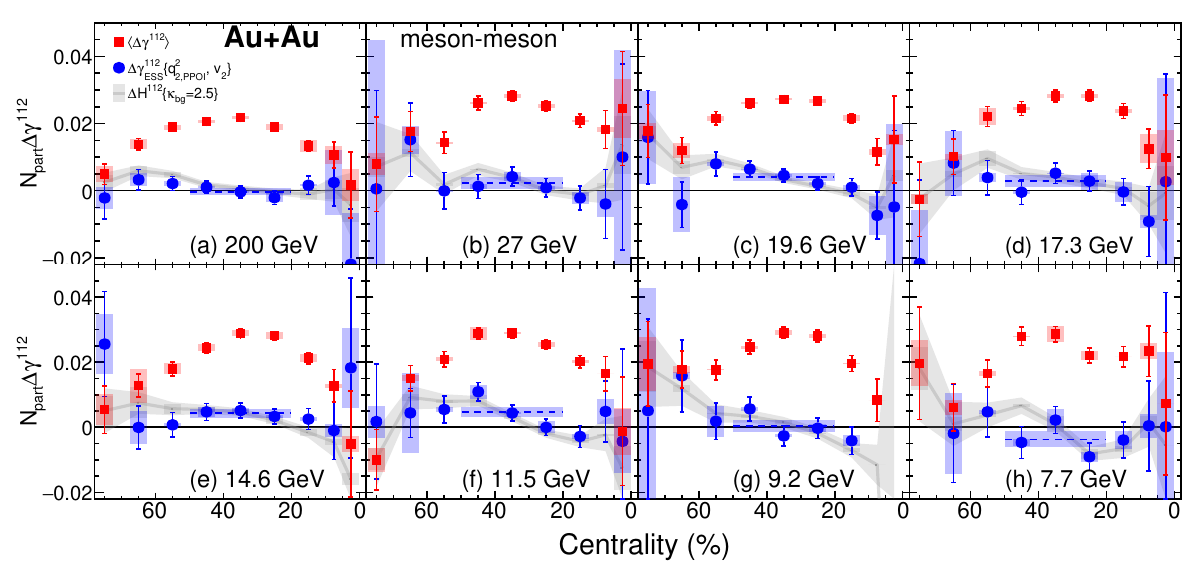}
\caption{Centrality dependence of 
$N_{\rm part}\Delta\gamma^{112}$ (red squares) and $N_{\rm part}\Delta\gamma^{112}_{\rm ESS}$ (blue circles) in Au+Au collisions at $\sqrt{s_{NN}}$ = 7.7--200 GeV. The error bars and shaded boxes represent the statistical and systematic uncertainties, respectively.
For comparison, $N_{\rm part}\Delta H^{112}(\kappa_{\rm{bg}}=2.5)$ is also shown with gray band denotes the statistic uncertainties.
The horizontal dashed lines denote constant fits to $N_{\rm part}\Delta\gamma^{112}_{\rm ESS}$ over the 20\%--50\% centrality region. }
\label{dg112-ESS}
\end{figure*}

\section{Results and discussion}
\label{measurements-star}
\subsection{Ensemble average}

In this subsection, we present the inclusive measurements of $v_2$, $\gamma^{112}_{\rm OS(SS)}$, $\gamma^{132}_{\rm OS(SS)}$, and $\delta_{\rm OS(SS)}$ averaged over the entire event ensemble.
Figure~\ref{v2-EPD} shows the centrality dependence of $v_2$ for charged mesons in Au+Au collisions using the event plane from the ZDC-SMD at 200 GeV (a) and the EPD at BES-II energies (b).
At each beam energy, the $v_2$ values exhibit a rising-then-falling trend. 
This feature contrasts with 
other analyses that use the participant plane or two-particle correlations to estimate $v_2$~\cite{v2_200GeV,STAR-11}.
In these analyses, $v_2$ keeps increasing or roughly saturates toward peripheral events because of the nonflow contributions.
For each bin in the 0\%--60\% centrality region, the magnitude of $v_2$ increases with rising beam energy, suggesting a stronger collective motion at higher energies.
In peripheral collisions (60\%--80\% centrality), the beam energy dependence of $v_2$ appears to transition to a reversed ordering
for all energies except for 7.7 and 9.2 GeV. 
The deviation of the two lowest energy points from the trend implies a shift in the underlying physics mechanism, such as the possible disappearance of the QGP.
Since nonflow contributions are minimized by using the spectator plane, the reversed $v_2$ ordering in 70\%--80\% collisions may be attributed to transported quarks~\cite{STAR-BESv1,STAR-BESv2} or electromagnetic field effects~\cite{STAR-Bfield,B_on_v2_1,B_on_v2_2,B_on_v2_3,B_on_v2_4}, which are beyond the scope of this work.

Figure~\ref{dg112} shows 
$\gamma^{112}_{\rm OS}$ and $\gamma^{112}_{\rm SS}$ as a function of centrality at all beam energies under study in this work.
The mean of $\gamma^{112}_{\rm OS}$ and $\gamma^{112}_{\rm SS}$ reveals the charge-independent background arising from momentum conservation and elliptic flow, displaying a pattern similar to the BES-I results~\cite{STAR-4}.
For all beam energies, the mid-central collisions produce an excess in $\gamma^{112}_{\rm OS}$ over $\gamma^{112}_{\rm SS}$, consistent with our expectations for the CME effect.
The aforementioned flow-related background will be discussed in the following subsection.

Figure~\ref{ddelta} is similar to Fig.~\ref{dg112}, but it displays the two-particle correlation observables $\delta_{\rm OS}$ and $\delta_{\rm SS}$. Note that momentum conservation generates a charge-independent background of order $-v_2/N$ for $\gamma^{112}$, but it contributes $-1/N$ to both $\delta_{\rm OS}$ and $\delta_{\rm SS}$~\cite{Flow_CME}.
In mid-central collisions, the excess of $\delta_{\rm OS}$ compared with $\delta_{\rm SS}$ contradicts the CME expectation~\cite{STAR-4}, 
indicating that there are important charge-dependent backgrounds for $\Delta\gamma^{112}$ in the centrality range 10--60\%. 

\begin{figure*}[!tbhp]
\includegraphics[width=\textwidth]{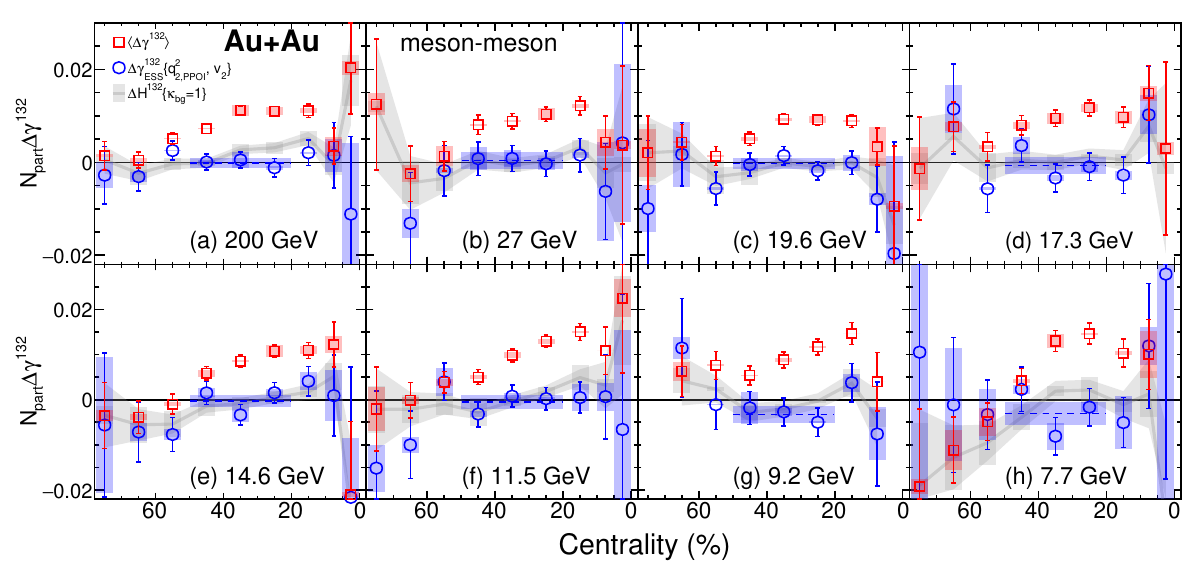}
\caption{Centrality dependence of 
$N_{\rm part}\Delta\gamma^{132}$ (red square) and $N_{\rm part}\Delta\gamma^{132}_{\rm ESS}$ (blue circle) in Au+Au collisions at $\sqrt{s_{NN}}$ = 7.7--200 GeV. The error bars and shaded boxes represent the statistical and systematic uncertainties, respectively.
For comparison, $N_{\rm part}\Delta H^{132}(\kappa_{\rm{bg}}=1)$ is also shown with gray band denotes the statistic uncertainties. 
The horizontal dashed lines denote constant fits to $N_{\rm part}\Delta\gamma^{132}_{\rm ESS}$ over the 20\%--50\% centrality region.}

\label{dg132-ESS}
\end{figure*}

\begin{figure*}[!tbhp]
\includegraphics[width=\textwidth]{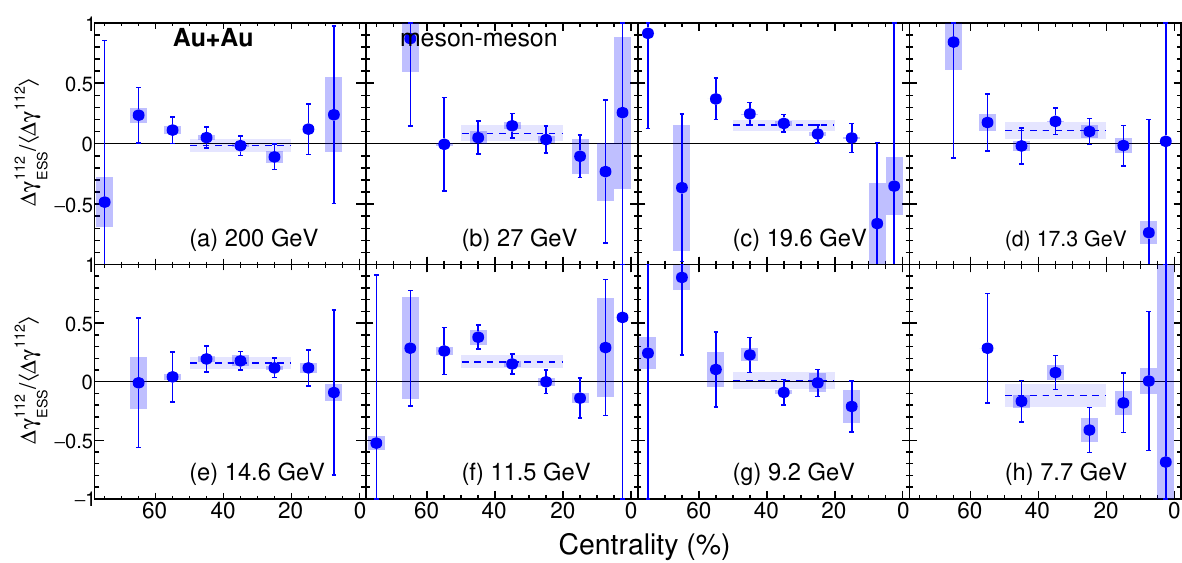}
\caption{Centrality dependence of 
the ratio $\gamma^{112}_{\rm ESS}/\gamma^{112}$ in Au+Au collisions at $\sqrt{s_{NN}}$ = 7.7--200 GeV. The error bars and shaded boxes represent the statistical and systematic uncertainties, respectively. The horizontal dashed lines denote constant fits over the 20\%--50\% centrality region.
}
\label{ratio-112}
\end{figure*}

Figure~\ref{dg132} is similar to Fig.~\ref{dg112}, except that it shows the $\gamma^{132}_{\rm OS}$ and $\gamma^{132}_{\rm SS}$ results. 
$\gamma^{132}_{\rm OS(SS)}$ is expected to be equal to $v_2\delta_{\rm OS(SS)}$~\cite{Subikash}, so we include $v_2\delta_{\rm OS}$ and $v_2\delta_{\rm SS}$ (shaded bands) for comparison.
In most cases, $\gamma^{132}_{\rm OS(SS)}$ follows 
$v_2\delta_{\rm OS(SS)}$ reasonably well, validating
$\gamma^{132}$ as a useful background indicator arising from the coupling between elliptic flow and other mechanisms.
At 27 and 9.2 GeV, $\gamma^{132}_{\rm OS}$ and $\gamma^{132}_{\rm SS}$ systematically fall below $v_2\delta_{\rm OS}$ and $v_2\delta_{\rm SS}$, respectively, indicating a residual momentum conservation effect due to differing performances of the forward and backward halves of the TPC.
Fortunately, the extra charge-independent background introduced by this effect will cancel out in $\Delta\gamma^{132}$ and will not affect the subsequent discussions. 
(See Sec. IV~C. for more discussions about this point.)

\subsection{Event shape selection}

Figures~\ref{dg112-ESS} and \ref{dg132-ESS} show the centrality dependence of $N_{\rm part}\Delta\gamma^{112}$ and $N_{\rm part}\Delta\gamma^{132}$, respectively, in Au+Au collisions at $\sqrt{s_{NN}}$= 7.7--200 GeV. 
The data points for the 0--5\% centrality interval at 9.2 GeV are unavailable due to poor event plane resolution.
For both observables, the ESS results (circles) exhibit a substantial reduction from the ensemble averages (squares).
Whereas $N_{\rm part}\Delta\gamma^{132}_{\rm ESS}$ aligns with zero in all centrality intervals at all beam energies, affirming the effectiveness of the ESS method, $N_{\rm part}\Delta\gamma^{112}_{\rm ESS}$ is finite in mid-central collisions at several energies from 11.5 to 19.6 GeV. 
For comparison, we also plot $N_{\rm part}\Delta H^{112}(\kappa_{\rm{bg}}^{112}=2.5)$ and $N_{\rm part}\Delta H^{132}(\kappa_{\rm{bg}}^{132}=1)$ in Figs.~\ref{dg112-ESS} and Fig.\ref{dg132-ESS} respectively. 
For each centrality interval at each beam energy, with $\kappa_{\rm{bg}}^{112(132)}$ set to 2.5 (1), the $N_{\rm part}\Delta H^{112(132)}$ result is consistent with the corresponding ESS value for $N_{\rm part}\Delta \gamma^{112(132)}$. The link between $\Delta H^{112(132)}$ and $\Delta \gamma_{\rm ESS}^{112(132)}$ will be further discussed in the next subsection.

We performed the constant fits to combine the ESS results over the 20\%--50\% centrality region as follows. $\Delta\gamma^{132}_{\rm{ESS}}$ is consistent with zero at all energies.
At 200 GeV,  $\Delta\gamma^{112}_{\rm{ESS}}$ is consistent with zero,
corroborating previous STAR measurements ~\cite{STAR-8,STAR-6}. 
While $\Delta\gamma^{112}_{\rm{ESS}}$ aligns with zero at  7.7 and 9.2 GeV, finite charge separation values are seen at 11.5, 14.6, and 19.6 GeV, with significance levels of 2.6$\sigma$, 3.1$\sigma$, and 3.3$\sigma$, respectively. 
The data at 17.3 and 27 GeV also show positive results but with a lower significance of 1.3$\sigma$ and 1.1$\sigma$, respectively.

\begin{figure}[tbhp]
\includegraphics[width=0.48\textwidth]{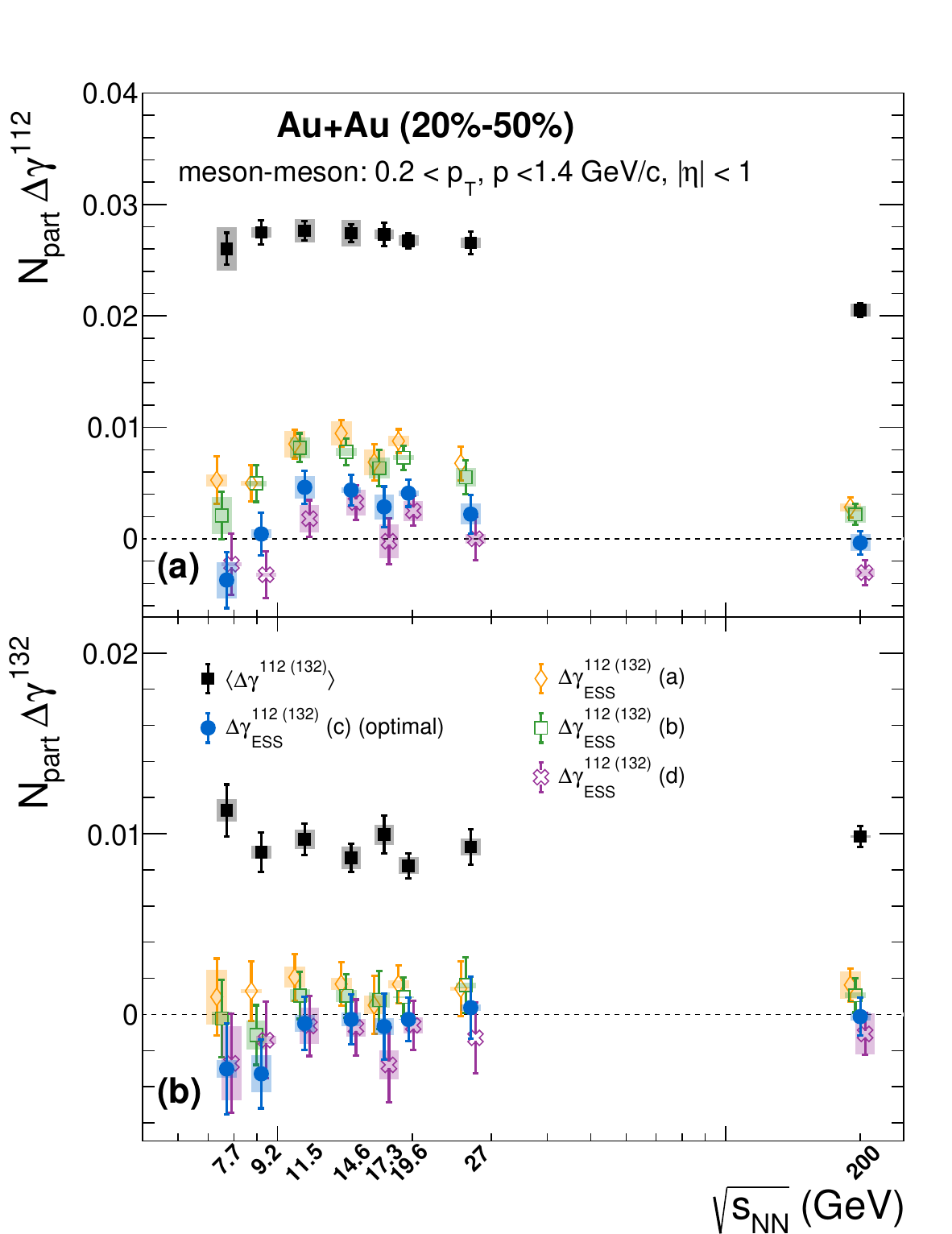}
\caption{Beam energy dependence of 
$N_{\rm part}\langle\Delta\gamma^{112}\rangle$ (a) and $N_{\rm part}\langle\Delta\gamma^{132}\rangle$ (b), together with ESS results obtained using four recipes, in 20--50\% Au+Au collisions. The error bars and shaded boxes represent the statistical and systematic uncertainties, respectively.
}
\label{bes-2-dg112}
\end{figure}

Figure~\ref{ratio-112} shows the centrality dependence of the ratio of $\Delta\gamma^{112}_{\rm{ESS}}$ to $\Delta\gamma^{112}$ in Au+Au collisions at $\sqrt{s_{NN}}$= 7.7--200 GeV. 
In the following discussions in this subsection, we will focus on the ratio in the 20\%--50\% centrality region obtained with constant fits, which we refer to as the fraction of the background-subtracted charge separation potentially associated with the CME, or $f_{\rm CME}$.
Generally, at least 80\% of $\Delta\gamma^{112}$ arises from the flow-related background.
At 200 GeV, the ratio ($f_{\rm CME}$) is $[-1.4 \pm 5.1 ({\rm stat.})\pm3.1 ({\rm syst.})]\%$, offering an upper limit of 10\% with a 95\% confidence level~\cite{stat-unify}, compatible with previous Au+Au data measurements by STAR~\cite{STAR-8,STAR-6}. Based on model studies~\cite{ESS-1}, this upper limit can be translated into a value of 5\% for isobar collisions at 200 GeV using participant planes. 
This indicates the difference in the charge separation observables between the two isobaric systems should not exceed 0.7\% with a 95\% confidence level, assuming the relative $|\vec{B}|^2$ difference is 14\% between the two isobaric systems. 
This explains why no predefined CME signatures were observed in the isobar data, where the statistical uncertainty for the difference is at best 0.4\%~\cite{Isobar-2}.

At $\sqrt{s_{NN}}$ = 27 GeV, $f_{\rm CME}$ is $[8.6\pm 6.5({\rm stat.})\pm 3.7({\rm syst.})]\%$, consistent with the upper limit estimated from a prior STAR measurement~\cite{STAR-27-CME-1}.
At  11.5, 14.6, 17.3 and 19.6 GeV, the finite ratios exhibit significance levels of $2.8\sigma$, $3\sigma$, $1.4\sigma$ and $3.2\sigma$, similar to those of $N_{part}\Delta\gamma^{112}_{\rm ESS}$.
Assuming comparable physics conditions between 10 and 20 GeV, the average of the ratios is $[15.9 \pm 2.7({\rm stat.}) \pm 1.1({\rm syst.})] \%$, yielding over $5\sigma$ significance of the charge separation signal in 20\%--50\% collisions. 
At 7.7 and 9.2 GeV, the ratios are consistent with zero within uncertainties, which corresponds to an upper limit of 11\% and 15\%, respectively, with a 95\% confidence level.

Figure ~\ref{bes-2-dg112} depicts the beam energy dependence of $N_{\rm part}\langle\Delta\gamma^{112(132)}\rangle$ and $N_{\rm part}\Delta\gamma^{112(132)}_{\rm ESS}$ obtained using four ESS recipes in 20--50\% in Au+Au collisions.
In general, both $\gamma^{112}_{\rm ESS}$ and $\Delta\gamma^{132}_{\rm ESS}$ follow the expected ordering among the four recipes at each energy, though sizable statistical uncertainties at lower energies may cause slight deviations from this ordering.
The $\Delta\gamma^{132}_{\rm ESS}$ values are consistent with zero across all recipes, but a $\chi^2$ test indicates that recipe (c) is the optimal choice, being closest to zero with the smallest $\chi^2$.
For simplicity, we denote $\Delta\gamma^{112}_{\rm ESS}$ as the result from recipe (c) in the following discussion. 
Although the ensemble average $N_{\rm part} \langle\Delta\gamma^{112}\rangle$ remains relatively constant, 
$N_{\rm part}\Delta\gamma^{112}_{\rm{ESS}}$ reveals a discernible and statistically significant charge separation signal
at $\sqrt{s_{NN}}$= 11.5, 14.6, and 19.6 GeV, with significance levels of 2.6$\sigma$, 3.1$\sigma$, and 3.3$\sigma$, respectively. 
The remaining charge separation signals are consistent with zero at $\sqrt{s_{NN}}$= 7.7, 9.2, and 200 GeV.
The significance levels of the positive $N_{\rm part}\Delta\gamma^{112}_{\rm{ESS}}$ values at $\sqrt{s_{NN}}$= 17.3 and 27 GeV are $1.3\sigma$ and $1.1\sigma$, respectively.

\begin{figure*}[!tbhp]
\centering
\includegraphics[width=\textwidth]{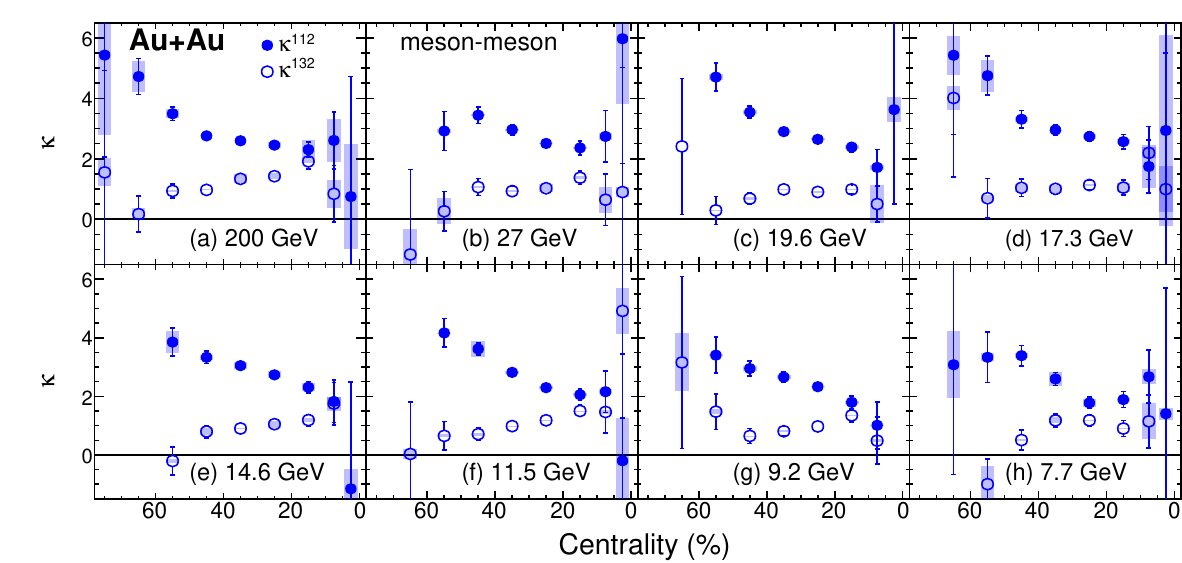}
\caption{Centrality dependence of the normalized correlator  $\kappa^{112}$ (solid circle) and $\kappa^{132}$ (open circle) in Au+Au collisions at $\sqrt{s_{NN}}$ = 7.7--200 GeV. 
}
\label{BES-kappa-inc}
\end{figure*}

\begin{figure*}[!tbhp]
\centering
\includegraphics[width=\textwidth]{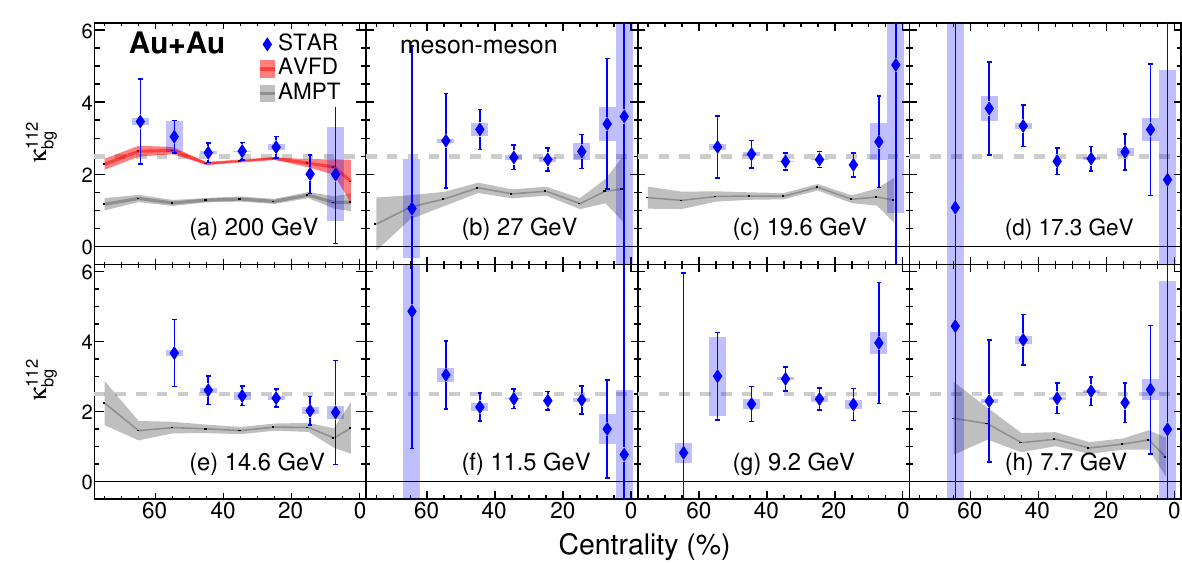}
\caption{Centrality dependence of the background coefficient  $\kappa_{\rm{bg}}^{112}$ in Au+Au collisions at $\sqrt{s_{NN}}$ = 7.7--200 GeV. The horizontal dashed line at 2.5 is to guide the eye.
For comparison, the simulation results from EBE-AVFD are shown at 200 GeV (red band) and the simulation results from AMPT are shown for 200, 27, 19.6, 14.6, and 7.7 GeV (gray band).
}
\label{BES-kappa-112}
\end{figure*}

\begin{figure*}[!tbhp]
\centering
\includegraphics[width=\textwidth]{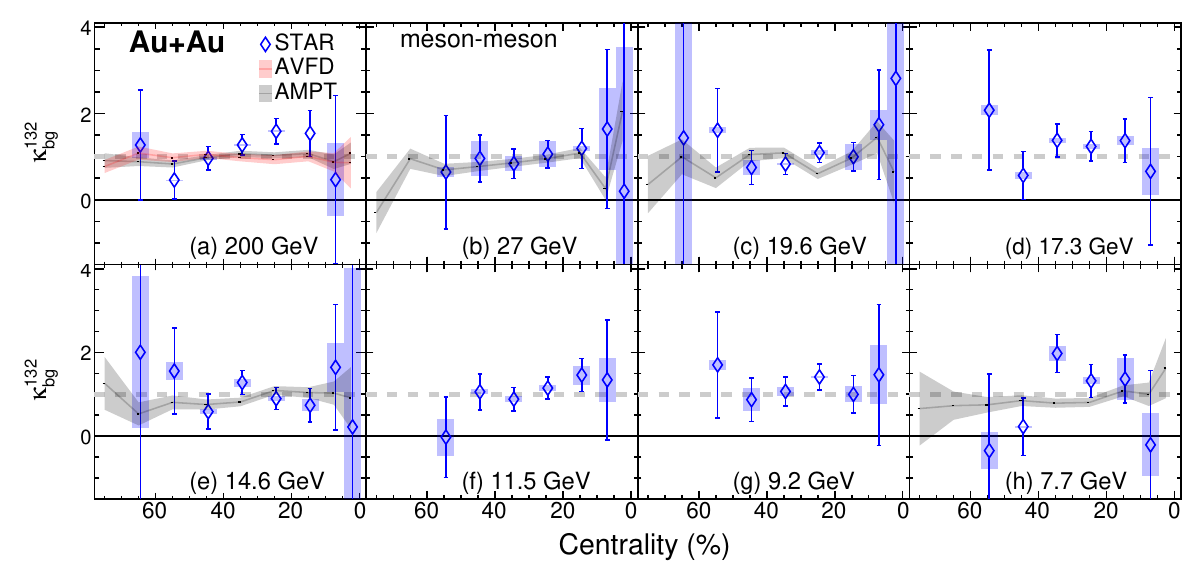}
\caption{Centrality dependence of the background coefficient  $\kappa_{\rm{bg}}^{132}$ in Au+Au collisions at $\sqrt{s_{NN}}$ = 7.7--200 GeV. The horizontal dashed line at 1 is to guide the eye. For comparison, the simulation results from EBE-AVFD are shown at 200 GeV (red band) and the simulation results from AMPT are shown for 200, 27, 19.6, 14.6, and 7.7 GeV (gray band).}
\label{BES-kappa-132}
\end{figure*}

 \begin{figure*}[tbhp]
\centering
\includegraphics[width=\textwidth]{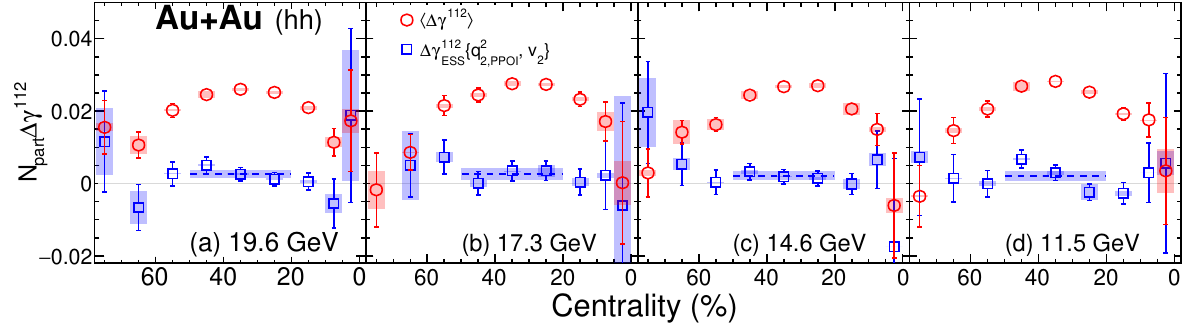}
\caption{Centrality dependence of 
$N_{\rm part}\Delta\gamma^{112}$ (red square) and $N_{\rm part}\Delta\gamma^{112}_{\rm ESS}$ (blue circle) in Au+Au collisions at $\sqrt{s_{NN}}$ = 19.6, 17.3, 14.6, and 11.5 GeV. The POI are all charged hadrons, including $p$ and ${\bar p}$. The horizontal dashed lines
denote constant fits to $N_{\rm part}\Delta\gamma^{112}_{\rm ESS}$ over the 20\%--50\% centrality region.}
\label{ESS-hh-fig}
\end{figure*}

\subsection{Flow-related background}

Next, we will investigate the flow-related background in $\Delta\gamma^{112(132)}$.
We will use the $H^{112(132)}$ correlator via the background coefficient $\kappa_{\rm{bg}}^{112(132)}$ because it quantifies the coupling between elliptic flow and the two-particle correlation. 
We first normalize the inclusive $\Delta\gamma^{112(132)}$ correlator with $v_2$ and $\Delta\delta$, as shown in Eq.(~\ref{kappa-equ}).
\begin{equation}
\kappa^{112(132)} \equiv \Delta\gamma^{112(132)}/(v_2\Delta\delta).  
\label{kappa-equ}
\end{equation}
The definition of $\kappa^{112(132)}$ assumes a factorization between $v_2$ and $\Delta\delta$, which can be verified with good precision. We examined the relation between $\Delta\delta$ and $v_2$ using the four ESS recipes (similar to Fig.~\ref{4ESS}, but with $\Delta\delta$ on the vertical axis). In all cases, $\Delta\delta$ remains approximately constant, with variations over the relevant $v_2$ range limited to at most 1\%.

Figure~\ref{BES-kappa-inc} shows $\kappa^{112}$ and $\kappa^{132}$ as a function of centrality in Au+Au collisions at 7.7--200 GeV. 
Note that $\kappa^{132}$ remains roughly constant around unity while $\kappa^{112}$ is larger and has the same shape, with respect to  centrality, at all beam energies.
Also note that $\kappa^{112}$ is dominated by the flow-related background, just like $\Delta\gamma^{112}$.

Assuming the ESS method removes all of the flow-related background, we replace $H^{112(132)}$ in Eq.~(\ref{h-corr}) with $\gamma^{112(132)}_{\rm ESS}$ and estimate the background coupling constants with
\begin{eqnarray}
\kappa_{\rm{bg}}^{112} &=& \frac{\Delta\gamma^{112}-\Delta\gamma_{\mathrm{ESS}}^{112}}{v_2 (\Delta\delta+\Delta\gamma_{\mathrm{ESS}}^{112})}, \\  
\kappa_{\rm{bg}}^{132} &=& \frac{\Delta\gamma^{132}-\Delta\gamma_{\mathrm{ESS}}^{132}}{v_2(\Delta\delta+\Delta\gamma_{\mathrm{ESS}}^{132})}. 
\end{eqnarray}

Figures~\ref{BES-kappa-112} and \ref{BES-kappa-132} present the centrality dependence of $\kappa_{\rm{bg}}^{112}$ and $\kappa_{\rm{bg}}^{132}$, respectively, in Au+Au collisions at $\sqrt{s_{NN}}$ = 7.7--200 GeV. 
In all centrality intervals and at all beam energies, the measured $\kappa_{\rm{bg}}^{112}$ and $\kappa_{\rm{bg}}^{132}$ are consistent with 2.5 and 1, respectively, as indicated by horizontal dashed lines. 
Both $\kappa_{\rm{bg}}^{112}$ and $\kappa_{\rm{bg}}^{132}$ appear unaffected by centrality and beam energy, standing in contrast to the charge separation signal shown in Figs.~\ref{dg112-ESS} and \ref{bes-2-dg112}, as well as the difference between Figs.~\ref{BES-kappa-inc} and \ref{BES-kappa-112}, which reveals a strong sensitivity to both. 
These background coupling constants impose a unique constraint on any model that seeks to capture the system's evolution in Au+Au collisions across a spectrum of temperatures and baryon chemical potentials.

Figure~\ref{BES-kappa-112} also shows the $\kappa_{\rm{bg}}^{112}$ values estimated from AMPT and EBE-AVFD via $\Delta\gamma^{112}/(v_2\Delta\delta)$ in the pure-background scenario (without the CME). 
AMPT predicts $\kappa^{112}_{\rm{bg}} \approx 1.3$ independent of centrality at 7.7, 19.6, 27, and 200 GeV~\cite{Subikash,Acta.Phys.Sin}, significantly lower than the data. 
On the other hand, EBE-AVFD calculates $\kappa_{\rm{bg}}^{112}$ to be around 2.4 for 20\%--50\% Au+Au collisions at 200 GeV. 
We speculate multiple reasons why EBE-AVFD reasonably captures the background mechanism in $\Delta\gamma^{112}$, whereas AMPT does not.
AMPT primarily reproduces the flow background caused by transverse momentum conservation (TMC), aligning well with the PHOBOS $v_2$ measurement~\cite{Subikash}, but inadequately accounts for the LCC effect.
In contrast, EBE-AVFD is specifically tuned to reflect the LCC effect based on experimental data.
This partially explains why the ESS method removes the flow-related background in the EBE-AVFD model but tends to over-subtract it in AMPT~\cite{ESS-2}. 
Another key distinction between the two models is that EBE-AVFD is based on hydrodynamics, while AMPT is a microscopic transport model. 
Although both models reproduce $v_2$ for final-state particles well, their distinct flow development schemes result in different behavior in when and how $v_2$ couples with other mechanisms, such as LCC.

Both models predict $\kappa^{132}_{\rm{bg}} \approx 1$ independent of centrality across all available beam energies~\cite{Subikash,ESS-1,Acta.Phys.Sin}, consistent with the data.
Remarkably, at 27 and 9.2 GeV, the 
$\kappa_{\rm{bg}}^{132}$ values from the data remain consistent with 1, despite $\gamma^{132}_{\rm OS}$ and $\gamma^{132}_{\rm SS}$ being offset from the corresponding $v_2\delta_{\rm OS}$ and $v_2\delta_{\rm SS}$. 
This indicates that charge-independent backgrounds are effectively canceled out in $\Delta\gamma^{132}$.
Although $\kappa_{\rm{bg}}^{112}$ and $\kappa_{\rm{bg}}^{132}$ bear different values, they both appear to be constant over centrality and collision energy, implying qualitatively consistent relative strength in background mechanisms in $\Delta\gamma^{112}$  and $\Delta\gamma^{132}$, each characterized by a universal coupling between elliptic flow and a two-particle correlation.
Thus, it is reasonable to consider the diminishing $\Delta\gamma^{132}_{\rm ESS}$ values as a necessary data-driven validation of the ESS method in removing such flow-related backgrounds.
However, we do not consider $\Delta\gamma^{132}$ as an exact proxy for the background component in $\Delta\gamma^{112}$, merely scaled by a constant factor. For example, at 200 GeV, AMPT describes well $\kappa^{132}_{\rm bg}$ but fails to reproduce $\kappa^{112}_{\rm bg}$, demonstrating distinct underlying background sources.

\subsection{Measurements including $p$ and $\bar{p}$}

We further measure the charge separation, including (anti)protons, at $\sqrt{s_{NN}}$ between 10 and 20 GeV to assess the extent of this contamination affecting the results.
Figure~\ref{ESS-hh-fig} presents the centrality dependence of $N_{\rm part}\Delta\gamma^{112}$ and $N_{\rm part}\Delta\gamma^{112}_{\rm ESS}$ with (anti)protons included in the POI selection for Au+Au collisions at $\sqrt{s_{NN}} =$ 11.5, 14.6, 17.3, and 19.6 GeV. As expected, the inclusion of (anti)protons leads to a reduction of both the inclusive and ESS values.
In the 20\%--50\% centrality interval, the significance of $N_{\rm part}\Delta\gamma_{\rm ESS}$ drops to 1.3$\sigma$, 1.4$\sigma$, 1.2$\sigma$, and 1.8$\sigma$, respectively, from 11.5 to 19.6 GeV.

\section{Summary}\label{result}

The search for the CME in heavy-ion collisions is challenging because of the background contributions related to elliptic flow and nonflow. Accordingly, in this analysis of RHIC data with Au+Au collisions, we apply the event shape selection method to suppress the flow-related background and estimate the magnetic field direction using the spectator information to minimize nonflow effects. In the ESS approach, we broaden the traditional definitions of event shape and elliptic flow variables to include single particles as well as particle pairs, which results in four distinct analysis recipes. 
Our results from the four recipes follow an ordering expected by analytical derivations and models such as EBE-AVFD and AMPT~\cite{ESS-2}.
We report the final ESS results using the optimal recipe with the event shape variable based on particle pairs and the elliptic flow variable derived from single particles.

After background suppression, we have the following findings about the remaining charge separation signal $\Delta\gamma^{112}_{\rm ESS}$.
At $\sqrt{s_{NN}}$ = 200 GeV, $\Delta\gamma^{112}_{\rm ESS}$ is consistent with zero across all centrality intervals under investigation, implying the smallness of any possible CME signal or residual background due to the methodology. 
This result also helps explain the absence of CME signatures in the isobar data.
At each of three center-of-mass energies, 11.5, 14.6, and 19.6 GeV, $\Delta\gamma^{112}_{\rm ESS}$ is positive and finite in the 20\%--50\% centrality range, with around $3\sigma$ significance. 
The results at 17.3 and 27 GeV also show positive values but with a lower significance of $1.3\sigma$ and $1.1\sigma$.
When the data between 10 and 20 GeV are combined, the significance rises to over $5\sigma$.
This suggests an intriguing scenario where the chirality imbalance in quarks and
the intense magnetic field coexist to trigger the CME in mid-central Au+Au collisions at $\sqrt{s_{NN}}$ between 10 and 20 GeV. This energy region is near the presumed critical point, with a higher likelihood of topological vacuum transitions~\cite{Kharzeev-critical}. 
The time-dependent dynamics of the magnetic field~\cite{LHui-Bfield} could also favor lower beam energies with longer duration times, while at higher energies, the magnetic field may substantially decrease before the QGP formation~\cite{DynamicalEvolution}.
At 7.7 and 9.2 GeV, $\Delta\gamma^{112}_{\rm ESS}$ is consistent with zero. The restoration of chiral symmetry and/or the dominance of partonic degrees of freedom may diminish at such collision energies~\cite{Mizher:2010zb}, causing the precondition for the quark chirality imbalance to vanish.

We further investigate the background contributions. In general, about 80\% or more of $\Delta\gamma^{112}$ originates from the flow-related background.
$\gamma^{132}_{\rm OS(SS)}$, as a background indicator, aligns with $v_2\delta_{\rm OS(SS)}$ in most cases, capturing the essence of the flow-related background mechanism.
After background subtraction, $\Delta\gamma^{132}_{\rm ESS}$
is consistent with zero in each centrality interval at each beam energy. 
Based on the ESS results, we estimate the coupling between elliptic flow and the two-particle correlation in $\Delta\gamma^{112}$ and $\Delta\gamma^{132}$.
Whereas the magnetic field and chirality imbalance are both expected to strongly depend on centrality and beam energy, 
the background coupling constants appear to be universal, $\kappa^{112}_{\rm{bg}}\approx2.5$ and $\kappa^{132}_{\rm{bg}}\approx1$.
Thus, $\Delta\gamma^{112}$ and $\Delta\gamma^{132}$ exhibit similar background mechanisms but with different coupling constants. When $\Delta\gamma^{132}_{\rm ESS}$ is reduced to zero, it is reasonable to assume that the flow-related background in $\Delta\gamma^{112}$ is also effectively removed by the ESS method.

The detection of charge separation signals between 10 and 20 GeV, coupled with their absence elsewhere at lower and higher energies, highlights the need for further theoretical insights.
More experimental opportunities with higher statistics in heavy-ion collisions at $\sqrt{s_{NN}}$ between 10 and 20 GeV are highly anticipated to confirm the BES-II data.
At these beam energies, assuming $f_{\rm CME} \approx 20\%$ for isobar collisions in the 20\%–50\% centrality range, the relative difference in $\Delta\gamma^{112}$ between $^{96}_{44}$Ru+$^{96}_{44}$Ru and $^{96}_{40}$Zr+$^{96}_{40}$Zr could be around 3\%, which is within the reach of experimental precision.
Applying the ESS technique to the LHC data at several collision energies would establish a thorough understanding of the beam energy dependence of the CME search.

\begin{acknowledgments}{
We thank the RHIC Operations Group and SDCC at BNL, the NERSC Center at LBNL, and the Open Science Grid consortium for providing resources and support.  This work was supported in part by the Office of Nuclear Physics within the U.S. DOE Office of Science, the U.S. National Science Foundation, National Natural Science Foundation of China, Chinese Academy of Science, the Ministry of Science and Technology of China and the Chinese Ministry of Education, NSTC Taipei, the National Research Foundation of Korea, Czech Science Foundation and Ministry of Education, Youth and Sports of the Czech Republic, Hungarian National Research, Development and Innovation Office, New National Excellency Programme of the Hungarian Ministry of Human Capacities, Department of Atomic Energy and Department of Science and Technology of the Government of India, the National Science Centre and WUT ID-UB of Poland, the Ministry of Science, Education and Sports of the Republic of Croatia, German Bundesministerium f\"ur Bildung, Wissenschaft, Forschung and Technologie (BMBF), Helmholtz Association, Ministry of Education, Culture, Sports, Science, and Technology (MEXT), Japan Society for the Promotion of Science (JSPS) and Agencia Nacional de Investigaci\'on y Desarrollo (ANID) of Chile.
}
\end{acknowledgments}

{}

\begin{thebibliography}{}

\bibitem{CME-1}
D. Kharzeev, R. D. Pisarski, and M. H. G. Tytgat, 
Phys. Rev. Lett. {\bf 81}, 512 (1998).

\bibitem{CME-2}
D. Kharzeev and R. D. Pisarski, 
Phys. Rev. D  {\bf 61}, 111901(R) (2000).

\bibitem{CME-3}
D. Kharzeev, 
Phys. Lett. B  {\bf 633}, 260 (2006).

\bibitem{Kharzeev:2024zzm}
D.~E.~Kharzeev, J.~Liao and P.~Tribedy, Int. J. Mod. Phys. E {\bf 33}, 2430007 (2024).

\bibitem{Baryogenesis}
D.E. Kharzeev and J. Liao, Nat Rev Phys {\bf 3}, 55 (2021).

\bibitem{CME-5}
D. E. Kharzeev, L. D. McLerran, and H. J. Warringa, 
Nucl. Phys. A {\bf 803}, 227 (2008).

\bibitem{B-field-1}
V. Skokov, A. Illarionov, and V. Toneev, 
Int. J. Mod. Phys. A {\bf 24}, 5925 (2009).

\bibitem{B-field-2}
W.-T. Deng and X.-G. Huang, 
Phys. Rev. C {\bf 85}, 044907 (2012).

\bibitem{STAR-Bfield}
M. I. Abdulhamid {\it et al.} [STAR Collaboration],
Phys. Rev. X {\bf 14}, 011028 (2024).

\bibitem{STAR-1}
B. I. Abelev {\it et al.} [STAR Collaboration], 
Phys. Rev. Lett. {\bf 103}, 251601 (2009).

\bibitem{STAR-2}
B. I. Abelev {\it et al.} [STAR Collaboration], 
Phys. Rev. C {\bf 81}, 054908 (2010).

\bibitem{STAR-3}
L. Adamczyk {\it et al.} [STAR Collaboration], 
Phys. Rev. C {\bf 88}, 064911 (2013).

\bibitem{STAR-4}
L. Adamczyk {\it et al.} [STAR Collaboration], 
Phys. Rev. Lett. {\bf 113}, 052302 (2014).

\bibitem{STAR-5}
L. Adamczyk {\it et al.} [STAR Collaboration], Phys. Rev. C {\bf 89}, 044908 (2014).

\bibitem{STAR-6}
J. Adam {\it et al.} [STAR Collaboration], 
Phys. Lett. B {\bf 798}, 134975 (2019).

\bibitem{STAR-7}
M. S. Abdallah {\it et al.} [STAR Collaboration], 
Phys. Rev. C {\bf 106}, 034908 (2022).

\bibitem{STAR-8}
M. S. Abdallah {\it et al.} [STAR Collaboration], 
Phys. Rev. Lett.  {\bf 128}, 092301 (2022).

\bibitem{Isobar-1}
M. S. Abdallah {et al.} [STAR Collaboration], 
Phys. Rev. C {\bf 105}, 014901 (2022).

\bibitem{Isobar-2}
M. S. Abdallah {et al.} [STAR Collaboration],
arXiv:2308.16846; arXiv:2310.13096

\bibitem{STAR-27-CME-1}
B. E. Aboona {\it et al.} [STAR Collaboration],
Phys. Lett. B {\bf 839}, 137779 (2023).



\bibitem{ALICE-1}
B. Abelev {\it et al.} [ALICE Collaboration], 
Phys. Rev. Lett. {\bf 110}, 012301 (2013).

\bibitem{CMS-1}
V. Khachatryan {\it et al.} [CMS Collaboration], 
Phys. Rev. Lett. {\bf 118}, 122301 (2017).

\bibitem{CMS-2}
A. M. Sirunyan {\it et al.} [CMS Collaboration], 
Phys. Rev. C {\bf 97}, 044912 (2018).

\bibitem{ALICE-2}
S. Acharya {\it et al.} [ALICE Collaboration], 
Phys. Lett. B {\bf 777}, 151 (2018).

 \bibitem{ALICE-3}
S. Acharya {\it et al.} [ALICE Collaboration], 
J. High Energy Phys. {\bf 09}, 160 (2020).

\bibitem{DynamicalEvolution}
L. Yan and X.-G. Huang, Phys. Rev. D {\bf 107}, 094028 (2023).

\bibitem{STAR-BESI-mini-review}
J. Chen, X. Dong, X. He, H. Huang, F. Liu {\it et al.}, NUCL SCI TECH {\bf 35}, 214 (2024).

\bibitem{Sergeiflow1}
A. M. Poskanzer, S. A. Voloshin Phys. Rev. C {\bf 58}, 1671 (1998).

\bibitem{hydro}
U.~Heinz and R.~Snellings,
Ann.\ Rev.\ Nucl.\ Part.\ Sci.\  {\bf 63}, 123 (2013).


\bibitem{CME-8}
S. A. Voloshin, 
Phys. Rev. C {\bf 70}, 057901 (2004).


\bibitem{CME-9}
S. Choudhury et al., 
Chin. Phys. C {\bf 46}, 014101 (2022).

\bibitem{Magdy}
N. Magdy, S. Shi, J. Liao, N. Ajitanand, R. A. Lacey, Phys.
Rev. C {\bf 97}, 061901 (2018).

\bibitem{Tang}
A. H. Tang, Chin. Phys. C {\bf 44}, 054101  (2020).

\bibitem{PrattSorren:2011} 
S.~Schlichting and S.~Pratt, 
Phys.\ Rev.\ C {\bf 83}, 014913 (2011).

\bibitem{LCC-1}
W.-Y. Wu {\it et al.}, Phys. Rev. C {\bf 107}, L031902 (2023).

\bibitem{Pratt2010} 
S.~Pratt, S.~Schlichting and S.~Gavin, 
Phys.\ Rev.\ C {\bf 84}, 024909 (2011).

\bibitem{Flow_CME}
A. Bzdak, V. Koch and J. Liao, Lect. Notes Phys. {\bf 871}, 503 (2013).

\bibitem{STAR-v2-first}
K. H. Ackermann {\it et al.} [STAR Collaboration], 
Phys. Rev. Lett. {\bf 86} 402 (2001).

\bibitem{isobar_proposal1}
W.-T. Deng, X.-G. Huang, G.-L. Ma, and G. Wang, Phys. Rev. C {\bf 94}, 041901 (2016).
\bibitem{isobar_proposal2}
V. Koch, S. Schlichting, V. Skokov, P. Sorensen, J. Thomas, S. Voloshin, G. Wang, and H.-U. Yee, Chin. Phys. C {\bf 41},
072001 (2017).
\bibitem{isobar_proposal3}
W.-T. Deng, X.-G. Huang, G.-L. Ma, and G. Wang, Phys. Rev. C {\bf 97}, 044901 (2018).

\bibitem{TwoPlane1}
H.j. Xu, J. Zhao, X. Wang, H. Li, Z.W. Lin, C. Shen, F. Wang, Chin. Phys. C {\bf 42}, 084103 (2018). 
\bibitem{TwoPlane2}
S.A. Voloshin, Phys. Rev. C {\bf 98}, 054911 (2018).

\bibitem{InvMass}
J. Zhao, H. Li, and F. Wang, Eur. Phys. J. C {\bf 79}(2), 168, 2019.


\bibitem{SergeiESE}
J. Schukraft, A. Timmins, S. A. Voloshin, Phys. Lett. B {\bf 719}, 394 (2013).

\bibitem{rho00_dg112}
D. Shen, J. Chen, A.H. Tang and G. Wang, Phys. Lett B {\bf 839}, 10 (2023).
\bibitem{rho00_dg112_cont}
Z. Wang, J. Chen, D. Shen, A.H. Tang and G. Wang, Phys. Rev. C {\bf 111}, 014910 (2025). 

\bibitem{rho00_v2}
Z.-T. Liang and X.-N. Wang, Phys. Lett. B {\bf 629}, 20 (2005).

\bibitem{Xu:2023wcy}
Z.~Xu [STARColo],
EPJ Web Conf. \textbf{296} (2024), 04010
doi:10.1051/epjconf/202429604010
[arXiv:2401.00317 [nucl-ex]].

\bibitem{ESS-2}
Z. Xu, B. Chan, G. Wang, A. Tang, H. Huang,	
Phys. Lett. B {\bf 848}, 138367 (2024)

\bibitem{1st-ESS}
F. Wen, J. Bryon, L. Wen, G. Wang, Chin. Phys. C {\bf 42}, 014001 (2018). 

\bibitem{ESS-1}
R. Milton, G. Wang, M. Sergeeva, S. Shi, J. Liao, and H. Huang. 
Phys. Rev. C {\bf 104}, 064906 (2021). 

\bibitem{berndt-2}
H. Petersen, B. M\"uller, 
Phys. Rev. C, {\bf 88}, 044918 (2013).



\bibitem{Subikash} S. Choudhury, G. Wang, W. B. He, Y. Hu and H. Z. Huang, Eur. Phys. J. C {\bf 80}, 383 (2020).

\bibitem{short-letter}
[STAR Collaboration], arXiv:2506.00275.



\bibitem{TPC-1}
M. Anderson et al., 
Nucl. Instrum. Meth. A {\bf 499}, 659 (2003).

\bibitem{tof}
W.J. Llope {\it et al.} [STAR Collaboration], 
Nucl. Instrum. Meth. A 
{\bf 661}, S110 (2012).

\bibitem{VPD} 
W. J. Llope {\it et al.} [STAR Collaboration], Nucl. Instrum. Meth. A {\bf 522}, 252 (2004).

\bibitem{EPD-1}
J. Adams {\it et al.} [STAR Collaboration], Nucl. Instrum. Meth. A {\bf 968}, 163970 (2020).

\bibitem{ZDC}
C. Adler {\it et al.}, Nucl. Instrum. Meth. A {\bf 470}, 488 (2001). 

\bibitem{BBC}
C. A. Whitten Jr. [STAR Collaboration], AIP Conf. Proc. {\bf 980}
390 (2008).

\bibitem{nSigma1}
H. Bichsel, Nucl. Instrum. Meth. A {\bf 562}, 154 (2006).
\bibitem{nSigma2}
Y. Xu {\it et al.}, Nucl. Instrum. Meth. A {\bf 614}, 28 (2010).

\bibitem{STAR-11}
L. Adamczyk {\it et al.}  [STAR Collaboration],
Phys. Rev. C {\bf 86}, 054908 (2012). 


\bibitem{BRAHMS-whitepaper} 
I. Arsene {\it et al.} [BRAHMS Collaboration], 
Nucl. Phys. A {\bf 757}, 1 (2005).
\bibitem{PHOBOS-whitepaper} 
B. B. Back {\it et al.} [PHOBOS Collaboration], 
Nucl. Phys. A {\bf 757}, 28 (2005).
\bibitem{STAR-whitepaper} 
J. Adams {\it et al.} [STAR Collaboration], Nucl. Phys. A {\bf 757}, 102 (2005).
\bibitem{PHENIX-whitepaper} 
K. Adcox {\it et al.} [PHENIX Collaboration], 
Nucl. Phys. A {\bf 757}, 184 (2005).

\bibitem{STAR-BESv2}
L. Adamczyk {\it et al.}  [STAR Collaboration], Phys. Rev. Lett. {\bf 110}, 142301 (2013).

\bibitem{STAR-BESv1}
L. Adamczyk {\it et al.}  [STAR Collaboration],
Phys. Rev. Lett. {\bf 112}, 162301 (2014).

\bibitem{v1_62GeV} 
J. Adams {\it et al.} [STAR Collaboration], Phys. Rev. C {\bf 73}, 034903 (2006).

\bibitem{E877:1997zjw}
J. Barrette {\it et al.} [E877 Collaboration], Phys. Rev. C {\bf 56}, 3254 (1997). 

\bibitem{rn-CMS}
V. Khachatryan {\it et al.} [CMS Collaboration], Phys. Rev. C {\bf 92}, 034911 (2015).

\bibitem{decor-ATLAS}
M. Aaboud {\it et al.} [ATLAS Collaboration], Eur. Phys. J. C {\bf 78}, 142 (2018).

\bibitem{decor-ATLAS2}
G. Aad {\it et al.} [ATLAS Collaboration], Phys. Rev. Lett. {\bf 126}, 122301 (2021).

\bibitem{Maowu}
M. Nie [STAR Collaboration], Nucl. Phys. A {\bf 1005}, 121783 (2021).

\bibitem{non-inter-dept}
Z. Xu, G. Wang, A. H. Tang, H. Z. Huang, 
Phys. Rev. C \textbf{107}, L061902 (2023).

\bibitem{Barlow-1}
R. Barlow, 
Conference on Advanced Statistical Techniques in Particle Physics, 2002, pp. 134–144. 

\bibitem{FlatPrior}
A. Zellner, An Introduction to Bayesian Inference in Econometrics, John Wiley \& Sons, New York, pp. 41–53 (1971). ISBN 0-471-98165-6.

\bibitem{Lyons}
L. Lyons, J. Phys. A: Math. Gen. {\bf 25}, 1967 (1992).

\bibitem{v2_200GeV}
J. Adams {\it et al.} [STAR Collaboration], Phys. Rev. C {\bf 72}, 14904 (2005).

\bibitem{B_on_v2_1}
R. K. Mohapatra, P. S. Saumia, and A. M. Srivastava, Mod. Phys. Lett. A {\bf 26}, 2477 (2011).

\bibitem{B_on_v2_2}
K. Tuchin, J. Phys. G: Nucl. Part. Phys. {\bf 39}, 025010 (2012).

\bibitem{B_on_v2_3}
S. I. Finazzo, R. Critelli, R. Rougemont, and J. Noronha, Phys. Rev. D {\bf 94}, 054020 (2016).

\bibitem{B_on_v2_4}
U. G\"ursoy, D. Kharzeev, E. Marcus, K. Rajagopal, and C. Shen, Phys. Rev. C {\bf 98}, 055201 (2018).

\bibitem{Acta.Phys.Sin}
Q. Y. Shou, J. Zhao, H. J. Xu, W. Li, G. Wang, A. H. Tang and F. Q. Wang, Acta Phys. Sin. 72, 112504 (2023).




\bibitem{stat-unify}
G. Feldman and R. Cousins,
Phys. Rev. D {\bf 57}, 3873 (1998).


\bibitem{Kharzeev-critical}
K. Ikeda, D. E. Kharzeev, and Y. Kikuchi, 
Phys. Rev. D {\bf 103}, L071502, (2021).


\bibitem{LHui-Bfield}
H. Li, X. Xia, X. G. Huang and H. Huang, Phys. Rev. C {\bf 108}, 044902 (2023).


\bibitem{Mizher:2010zb}
A. J. Mizher, M. N. Chernodub, and E. S. Fraga, Phys. Rev. D {\bf 82}, 105016 (2010).

\bibitem{par-p-ratio}
L. Adamczyk {\it et al.} [STAR Collaboration], Phys. Rev. C {\bf 96}, 044904 (2017).
































\end{thebibliography}
\end{document}